\documentclass[journal]{IEEEtran} 

\usepackage{amsmath,amssymb}
\usepackage{subfigure}
\usepackage{url,booktabs}
\usepackage{graphicx}
\usepackage{balance}
\usepackage{multirow}
\usepackage{cite}
\usepackage{algpseudocode}
\usepackage{units}

\newtheorem{LEO}{LEO}

\newtheorem{PEO}{PEO}

\newtheorem{Lemma}[LEO]{Lemma}
\newtheorem{Proposition}[PEO]{Proposition}

\begin{document}

\title{Entropy-based Statistical Analysis of PolSAR Data}

\author{Alejandro C.\ Frery,~\IEEEmembership{Member}, 
Renato J.\  Cintra,~\IEEEmembership{Senior Member}, and
Abra\~ao D.\ C.\ Nascimento,~\IEEEmembership{Student Member}
\thanks{This work was supported by CNPq, Fapeal and FACEPE, Brazil.}
\thanks{
A.\ C.\ Frery is with the Instituto de Computa\c c\~ao, Universidade Federal de Alagoas, BR 104 Norte km 97, 57072-970, Macei\'o, AL, Brazil, email: acfrery@gmail.com}
\thanks{R.\ J.\ Cintra and A.\ D.\ C.\ Nascimento  are with the Departamento de Estat\'istica, Universidade Federal de Pernambuco, Cidade Universit\'aria, 50740-540, Recife, PE, Brazil, e-mail: rjdsc@stat.ufpe.org, abraao.susej@gmail.com}
}

\markboth{IEEE Transactions on Geoscience and Remote Sensing}
{Nascimento \MakeLowercase{\textit{et al.}}: Entropy}

\maketitle

\begin{abstract}

Images obtained from coherent illumination processes are contaminated with speckle noise, with polarimetric synthetic aperture radar (PolSAR) imagery as a prominent example.
With an adequacy widely attested in the literature, the scaled complex Wishart distribution is an acceptable model for PolSAR data.
In this perspective, we derive analytic expressions for the Shannon, R\'enyi, and restricted Tsallis entropies under this model.
Relationships between the derived measures and the parameters of the scaled Wishart law (i.e., the equivalent number of looks and the covariance matrix) are discussed.
In addition, we obtain the asymptotic variances of the Shannon and R\'enyi entropies when replacing distribution parameters by maximum likelihood estimators.
As a consequence, confidence intervals based on these two entropies are also derived and proposed as new ways of capturing contrast.
New hypothesis tests are additionally proposed using these results, and their performance is assessed using simulated and real data.
In general terms, the test based on the Shannon entropy outperforms those based on R\'enyi's.
\end{abstract}

\section{Introduction}\label{entropy:intro}

\IEEEPARstart{P}{olarimetric} synthetic aperture radar (PolSAR) has been used to describe earth surface phenomena~\cite{LeePottier2009PolarimetricRadarImaging}.
This technology uses coherent illumination which causes the interference pattern called `speckle'~\cite{UlabyElachi1990}, which is multiplicative by nature and, in the format here considered, is non-Gaussian.
This fact precludes the use of conventional tools in PolSAR image analysis, requiring specialized techniques.

The scaled complex Wishart distribution has been successfully employed as a statistical model for homogeneous regions in PolSAR images.
This law is at the core of segmentation~\cite{BeaulieuTouzi2004}, classification~\cite{KerstenandLeeandAinsworth2005}, and boundary detection~\cite{Schou2003} techniques.

The concepts of ``information'' and ``entropy'' were given formal mathematical definitions in the context of data communications by Shannon in 1948~\cite{Shannon1948}.
Thenceforth, the proposition and application of information and entropy measures have become an active research field in several areas.
Zografos and Nadarajah derived closed expressions for Shannon and R\'enyi entropies for several univariate~\cite{Nadarajah:2003}, bivariate~\cite{Nadarajah2005173}, and multivariate~\cite{Zografos200571} distributions.

Exploring relationships associated with the log-likelihood function, Zong~\cite{Song2011} applied the R\'enyi entropy to several univariate distributions.
In fact, this entropy measure has been applied to image processing problems such as data mining, detection, segmentation, and classification~\cite{Basseville1989,Maybank2007,Liunietal}. 
Another prominent entropy measure is the restricted Tsallis entropy. 
This tool was introduced by Tsallis in~\cite{Tsallis1998,Tsallis2002} and is related to the R\'enyi entropy. 
The restricted Tsallis entropy has found applications in statistical physics~\cite{Wilk20084809}.

Among these information theoretical tools, the Shannon entropy has been applied to PolSAR imagery.
Morio~\textit{et al.}~\cite{MorioRefregierGoudailFernandezDupuis2009} analyzed such entropy for the characterization of polarimetric targets under the complex, circular, and multidimensional Gaussian distribution.

Stochastic distances are also derived within the framework of information theory.
A comprehensive examination of these measures is presented and applied to intensity SAR data in~\cite{ HypothesisTestingSpeckledDataStochasticDistances,ParametricNonparametricTestsSpeckledImagery}, and to PolSAR models in~\cite{FreryNascimentoCintraChileanJournalStatistics2011}.

In this paper, we derive analytic expressions for the Shannon, R\'enyi, and restricted Tsallis entropies under the scaled complex Wishart law.
These measures are analyzed as particular cases of the ($h,\phi$)-entropy proposed by Salicr\'u~\textit{et al.}~\cite{salicruetal1993}.
When parameters are replaced by maximum likelihood estimators, these entropies become random variables; expressions for the asymptotic variances of the Shannon and R\'enyi entropies are derived (the Tsallis entropy becomes analytically intractable and, thus, is no further considered).

Novel methodologies for testing hypotheses and constructing confidence intervals are proposed for quantifying contrast in PolSAR imagery using these results. 
Such measures can be used in PolSAR segmentation~\cite{BeaulieuTouzi2004}, classification~\cite{KerstenandLeeandAinsworth2005}, boundary detection~\cite{PolarimetricSegmentationBSplinesMSSP,GambiniandMejailandJacobo-BerllesandFrery}, and change detection~\cite{IngladaMercier2007}.
Monte Carlo experiments are performed for assessing the performance of the discussed measures in synthetic data, and an application to real PolSAR data is performed.

The remainder of this paper is organized as follows.
Section~\ref{entropy:model} recalls the scaled Wishart law.
Selected information theoretic tools are summarized in Section~\ref{entropy:IT}.
Section~\ref{entropy:results} presents the proposed entropies and the asymptotic variance of their estimators, along with an application.
Section~\ref{entropy:conclusion} concludes the paper.

\section{The Complex Wishart distribution}\label{entropy:model}

Full-polarimetric SAR sensors record the complex scattering coefficient for the four combinations of the received and transmitted complex linear polarizations: $S_\text{HH}$ (horizontal-horizontal), $S_\text{HV}$ (horizontal-vertical), $S_\text{VH}$ (vertical-horizontal), and $S_\text{VV}$ (vertical-vertical).
When natural targets are considered, the conditions of the reciprocity theorem~\cite{UlabyElachi1990,Conradsen2003,LopezMartinezFabregas2003} are satisfied and it can be assumed that $S_\text{HV}=S_\text{VH}$.

In general, we may consider systems with $m$ polarization elements, which constitute a complex random vector denoted by:
\begin{equation}
\boldsymbol{y}=[S_1\; S_2\;\cdots\; S_m]^{t},
\label{backscattervectorr}
\end{equation}
where $(\cdot)^t$ is the transposition operator.
It is commonly assumed that the scattering vector $\boldsymbol{y}$ follows a circular complex Gaussian law~\cite{Goudail2004}.

Multilook PolSAR data are usually formed in order to enhance the signal-to-noise ratio~(SNR):
$$
\boldsymbol{Z}=\frac{1}{L}\displaystyle \sum_{i=1}^L \boldsymbol{y}_i \boldsymbol{y}_i^{\text{H}} ,
$$
where $(\cdot)^\text{H}$ is the Hermitian operator, $\boldsymbol{y}_i$ represents the scattering vector in the \textit{$i$}th look, and $L$ is the number of looks, a parameter related to the noise effect in SAR imagery. 
Matrix $\boldsymbol{Z}$ is defined over the set of positive-definite Hermitian matrices $\mathcal A$. 
Moreover, Goodman showed that $L\boldsymbol{Z}$ follows the ordinary complex Wishart law~\cite{Goodmanb} and, therefore, the density of the scaled random matrix $\boldsymbol{Z}$ is
\begin{align}
f_{\boldsymbol{Z}}(\boldsymbol{Z}';\boldsymbol{\Sigma},L)=\frac{L^{mL}|\boldsymbol{Z}'|^{L-m}}{|\boldsymbol{\Sigma}|^L \Gamma_m(L)}\exp\big[-L\operatorname{tr}(\boldsymbol{\Sigma}^{-1}\boldsymbol{Z}')\big],
\label{density1}
\end{align}
where $m$ is the order of $\boldsymbol{\Sigma}$, $m\leq L$, $\Gamma_m(L)=\pi^{m(m-1)/2}\prod_{k=0}^{m-1}\Gamma(L-k)$, $\Gamma(\cdot)$ is the gamma function, $\boldsymbol{\Sigma}=\operatorname{E}\{\boldsymbol{y}\boldsymbol{y}^{\text{H}}\}$, and $\operatorname{E}\{\cdot\}$ is the expectation operator.
This is denoted as $\boldsymbol{Z} \sim \mathcal W_m(\boldsymbol{\Sigma}, L)$.

\section{Information Theory}\label{entropy:IT}

Information theory provides important tools for statistical inference~\cite{BlattandHero}, data compression~\cite{Donohoetal1998}, and image processing~\cite{MorioRefregierGoudailFernandezDupuis2009}, to name a few applications.
In particular, entropy is a fundamental concept related to the notion of disorder in mechanical statistics~\cite{KullbackInformationTheoryStatistics}.
Salicr\'u~\textit{et al.}~\cite{salicruetal1993} proposed the ($h,\phi$)-entropy class, which generalizes the original concept.
In the following, we recall this definition and we derive entropies for positive-definite Hermitian random matrices.

Let $f_{\boldsymbol{Z}}(\boldsymbol{Z}';\boldsymbol{\theta})$ be a probability density function with parameter vector $\boldsymbol{\theta}$ which characterizes the distribution of the random matrix $\boldsymbol{Z}$.
The ($h,\phi$)-entropy relative to $\boldsymbol{Z}$ is defined by 
\begin{align*}
H_{\phi}^h(\boldsymbol{\theta})=h\Big(\int_{\mathcal A}\phi(f_{\boldsymbol{Z}}(\boldsymbol{Z}';\boldsymbol{\theta}))\mathrm{d}\boldsymbol{Z}'\Big),
\end{align*}
where either $\phi\colon\bigl[0,\infty\bigr) \rightarrow \mathbb{R}$ is concave and $h\colon\mathbb{R} \rightarrow \mathbb{R}$ is increasing, or $\phi$ is convex and $h$ is decreasing.   
The differential element $\mathrm{d}\boldsymbol{Z}'$ is given by
$$
\mathrm{d}\boldsymbol{Z}'=\prod_{i=1}^m\mathrm{d}Z_{ii}\prod^m_{\underbrace{i,j=1}_{i<j}}\mathrm{d}\Re\{Z_{ij}\} \mathrm{d}\Im\{Z_{ij}\},
$$ 
where $Z_{ij}$ is the $(i,j)$-th entry of matrix $\boldsymbol{Z}'$, and $\Re$ and $\Im$ denote the real and imaginary parts, respectively~\cite{Goodmanb}.
Table~\ref{entrpyexpression} shows the specification of $h$ and $\phi$ for the three entropies we use in this article: Shannon, R\'enyi, and restricted Tsallis.

\begin{table}[hbt]
\centering   
\caption{($h,\phi$)-entropies and related functions}
\begin{tabular}{ccc}\toprule 
{ $(h,\phi)$-{entropy}} & { $h(y)$} & { $\phi(x)$} \\ 
\cmidrule(lr{.25em}){1-1} \cmidrule(lr{.25em}){2-2} \cmidrule(lr{.25em}){3-3} 
Shannon~\cite{salicruetal1993} & $y$ & $-x\ln x$ \\
Restricted Tsallis (order $\beta \in \mathbb{R}_{+}\,:\,\beta\neq 1$)~\cite{Havrda1967} & $y$ & $\frac{x^\beta-x}{1-\beta} $ \\
R\'{e}nyi (order $\beta \in \mathbb{R}_+\,:\,\beta\neq 1$)~\cite{Renyi1961} & $\frac{\ln y}{1-\beta}$ & $x^\beta$  \\
\bottomrule                      
\end{tabular}
\label{entrpyexpression}
\end{table}

The following result, derived by Pardo~\textit{et al.}~\cite{Pardo1997}, paves the way for the proposal of asymptotic statistical inference methods based on entropy.

\begin{Lemma}\label{col1}
Let $\widehat{\boldsymbol{\theta}}=[\widehat{\theta_1}\;\widehat{\theta_2}\;\cdots\;\widehat{\theta_p}]^t$ be the ML estimate of the parameter vector $\boldsymbol{\theta}=[\theta_1\;\theta_2\;\cdots\;\theta_p]^t$ based on a random sample of size $N$ under the model $f(\boldsymbol{Z}';\boldsymbol{\theta})$. 
Then
$$
\sqrt{N} \big[H_h^\phi(\widehat{\boldsymbol{\theta}})-H_h^\phi(\boldsymbol{\theta})\big] \xrightarrow[N\rightarrow \infty]{\mathcal D} \mathcal N(0,\sigma_H^2(\boldsymbol{\theta})),
$$
where $\mathcal N(\mu,\sigma^2)$ is the Gaussian distribution with mean $\mu$ and variance $\sigma^2$, `$\xrightarrow[]{\mathcal{D}}$' denotes convergence in distribution, 
\begin{equation}
\sigma_H^2(\boldsymbol{\theta})=\boldsymbol{\delta}^t \mathcal K(\boldsymbol{\theta})^{-1}\boldsymbol{\delta},
\label{varentro}
\end{equation}
$\mathcal K(\boldsymbol{\theta})=\operatorname{E}\{-\partial^2 \ln f_{\boldsymbol{Z}}(\boldsymbol{Z};\boldsymbol{\theta})/\partial \boldsymbol{\theta}^2\}$ is the Fisher information matrix, and $\boldsymbol{\delta}=[\delta_1\;\delta_2\;\cdots\;\delta_p]^t$ such that 
$\delta_i=\partial H_h^\phi(\boldsymbol{\theta})/\partial \theta_i$ for $i=1,2,\ldots,p$.
\end{Lemma}

In the following we introduce a methodology for hypothesis tests and confidence intervals based on entropy.

\subsection{Hypothesis test}

Let $\boldsymbol{Z}_i$ be a positive-definite random matrix with probability density function defined over $\boldsymbol{\mathcal A}$ with parameter vector $\boldsymbol{\theta}_i$ for $i=1,2,\ldots,r$, where $r$ is the number of populations to be assessed.
We are interested in testing the following hypotheses: 
\begin{equation*}
\left\{
\begin{array}{r}
\mathcal H_0 \colon H_h^\phi(\boldsymbol{\theta}_1)=H_h^\phi(\boldsymbol{\theta}_2)=\cdots=H_h^\phi(\boldsymbol{\theta}_r)=v,\\
\mathcal H_1 \colon H_h^\phi(\boldsymbol{\theta}_i) \neq H_h^\phi(\boldsymbol{\theta}_j)\text{ for some }  i \text{ and } j.
\end{array}
\right.
\end{equation*}
In other words, statistical evidence is sought for assessing whether at least one of the $r$ regions of a PolSAR image has different entropy when compared to the remaining regions.

Let $\widehat{\boldsymbol{\theta}_i}$ be the ML estimate for $\boldsymbol{\theta}_i$ based on a random sample of size $N_i$ under $\boldsymbol{Z}_i$, for $i=1,2,\ldots,r$.  
From Lemma~\ref{col1} we have that
$$
\frac{\sqrt{N_i}\big(H_h^\phi(\widehat{\boldsymbol{\theta}_i})-v\big)}{\sigma_H(\widehat{\boldsymbol{\theta}_i})} \xrightarrow[N_i\rightarrow \infty]{\mathcal D} \mathcal N(0,1)
$$ 
for $i=1,2,\ldots,r$.
Therefore,
\begin{equation}
\sum_{i=1}^r\frac{N_i\big(H_h^\phi(\widehat{\boldsymbol{\theta}_i})-v\big)^2}{\sigma_H^2(\widehat{\boldsymbol{\theta}_i})}\xrightarrow[N_i\rightarrow \infty]{\mathcal D} \chi^2_r.
\label{cocran0}
\end{equation}
Since $v$ is, in practice, unknown, in the following we modify this test statistic in order to take this into account.
Considering an application of Cochran's theorem~\cite{CambridgeJournals:2026696}, we obtain:
\begin{align}
\sum_{i=1}^r&\frac{N_i\big(H_h^\phi(\widehat{\boldsymbol{\theta}_i})-v\big)^2}{\sigma_H^2(\widehat{\boldsymbol{\theta}_i})}
=\nonumber\\
&\sum_{i=1}^r\frac{N_i\big(H_h^\phi(\widehat{\boldsymbol{\theta}_i})-\overline{v}\big)^2}{\sigma_H^2(\widehat{\boldsymbol{\theta}_i})}
+\sum_{i=1}^r\frac{N_i\big(\overline{v}-v\big)^2}{\sigma_H^2(\widehat{\boldsymbol{\theta}_i})},
\label{cocran1}
\end{align}
where 
$$
\overline{v}=\bigg[\sum_{i=1}^r\frac{N_i}{\sigma_H^2(\widehat{\boldsymbol{\theta}_i})}\bigg]^{-1}\sum_{i=1}^r \frac{N_iH_h^\phi(\widehat{\boldsymbol{\theta}_i})}{\sigma_H^2(\widehat{\boldsymbol{\theta}_i})}.
$$

Salicr\'u~\textit{et al.}~\cite{salicruetal1993} showed that the second summation in the right-hand side of Equation~\eqref{cocran1} is chi-square distributed with one degree of freedom.
Since the left-hand side of~\eqref{cocran1} is chi-square distributed with $r$ degrees of freedom~(cf. Equation~\eqref{cocran0}), we conclude that:
$$
\sum_{i=1}^r\frac{N_i\big(H_h^\phi(\widehat{\boldsymbol{\theta}_i})-\overline{v}\big)^2}{\sigma_H^2(\widehat{\boldsymbol{\theta}_i})}\xrightarrow[N_i\rightarrow \infty]{\mathcal D} \chi^2_{r-1}.
$$
In particular, consider the following test statistic:
\begin{align}\label{teststatisticonentropy}
S_{\phi}^h(\widehat{\boldsymbol{\theta}_1},\widehat{\boldsymbol{\theta}_2},\ldots,\widehat{\boldsymbol{\theta}_r})=\sum_{i=1}^r\frac{N_i\big(H_h^\phi(\widehat{\boldsymbol{\theta}_i})-\overline{v}\big)^2}{\sigma_H^2(\widehat{\boldsymbol{\theta}_i})}.
\end{align} 
We are now in the position to state the following result.

\begin{Proposition}\label{p1}
Let $N_i$, $i=1,2,\ldots,r$, be sufficiently large.
If $S_{\phi}^h(\widehat{\boldsymbol{\theta}_1},\widehat{\boldsymbol{\theta}_2},\ldots,\widehat{\boldsymbol{\theta}_r})=s$, then the null hypothesis $\mathcal H_0$ can be rejected at a level $\alpha$ if $\Pr\bigl( \chi^2_{r-1}>s\bigr)\leq \alpha$.
\end{Proposition}

\subsection{Confidence intervals}

Let $\widehat{\boldsymbol{\theta}}$ be the ML estimate of $\boldsymbol{\theta}$ for a sufficiently large sample $N$.
An approximate confidence interval for $H_h^\phi(\boldsymbol{\theta})$ at nominal level $\alpha$ is
\begin{equation}
H_h^\phi(\widehat{\boldsymbol{\theta}})\pm z_{\alpha/2} \sqrt{\frac{\sigma_H^2(\widehat{\boldsymbol{\theta}})}{N}},
\label{intervalforone}
\end{equation}
where $z_{\alpha/2}$ is the $\alpha/2$ quantile of the standard Gaussian distribution.

Consider now $\widehat{\boldsymbol{\theta}_1}$ and $\widehat{\boldsymbol{\theta}_2}$, ML estimates based on large samples $N_1$ and $N_2$, respectively.
An approximate confidence interval for $H_h^\phi(\widehat{\boldsymbol{\theta_1}})-H_h^\phi(\widehat{\boldsymbol{\theta_2}})$ is given by~\cite{Pardo1997}
$$
\bigl[H_h^\phi(\widehat{\boldsymbol{\theta}_1})-H_h^\phi(\widehat{\boldsymbol{\theta}_2}) \bigr]
\pm z_{\alpha/2} \sqrt{ \frac{\sigma_H^2(\widehat{\boldsymbol{\theta}_1})}{N_1}+\frac{\sigma_H^2(\widehat{\boldsymbol{\theta}_2})}{N_2}}.
$$

\section{Results}\label{entropy:results}

In the following, we derive results for the Shannon, R\'enyi, and restricted Tsallis entropies, denoted as $H_\text{S}$, $H^{\beta}_\text{R}$, and $H^{\beta}_\text{T}$, respectively, under the scaled complex Wishart law.
In particular, these measures are algebraically expressed, numerically evaluated, and assessed.
We adopt $\boldsymbol{\theta}=[L,\operatorname{vec}(\boldsymbol{\Sigma})^t]^t$, where $\operatorname{vec}(\cdot)$ is the vectorisation operator, as the working parameter vector.
Additionally, asymptotic results for these measures are computed.
To that end, we derive analytic expressions for the variance of the considered entropies. 
The entropies were evaluated at the ML estimate values.
Subsequently, hypothesis tests and confidence intervals are proposed based on Shannon and R\'enyi entropies.

\subsection{Expressions for the complex Wishart distribution}\label{entropy:results1}

We now present three schemes concerning the derivations of Shannon, restricted Tsallis, and R\'enyi entropies.

\subsubsection{Shannon entropy} 

Using the expression of the Shannon entropy obtained applying $h$ and $\phi$ from Table~\ref{entrpyexpression} to the density given in Equation~\eqref{density1}, we have that
\begin{align*}
H_{\text{S}}(\boldsymbol{\theta})&=-\int_{\boldsymbol{\mathcal{A}}} f_{\boldsymbol{Z}}(\boldsymbol{Z}';\boldsymbol{\Sigma},L)\ln f_{\boldsymbol{Z}}(\boldsymbol{Z}';\boldsymbol{\Sigma},L) \mathrm{d}\boldsymbol{Z}'.\\
&=\operatorname{E}\{-\ln f_{\boldsymbol{Z}}(\boldsymbol{Z})\}.
\end{align*}
Minor manipulations yield the following result: 
\begin{align*}
H_\text{S}(\boldsymbol{\theta})=&-mL\ln L + (m-L)\operatorname{E}\{\ln|\boldsymbol{Z}|\} \\
&+L\ln |\boldsymbol{\Sigma}| +\frac{m(m-1)}{2}\ln \pi+ \sum_{k=0}^{m-1}\ln \Gamma(L-k) \\
&+L\operatorname{E}\{\operatorname{tr}(\boldsymbol{\Sigma}^{-1}\boldsymbol{Z})\}.
\end{align*}

In~\cite{EstimationEquivalentNumberLooksSAR}, Anfinsen~\textit{et al.} obtain the following identity:
$$
\operatorname{E}\{\ln |\boldsymbol{Z}|\}=\ln|\boldsymbol{\Sigma}|+\psi_m^{(0)}(L)-m\ln L,
$$
where $\psi_m^{(0)}(\cdot)$ is the term of order zero of the $v$th-order multivariate polygamma function given by
$$
\psi_m^{(v)}(L)=\sum_{i=0}^{m-1} \psi^{(v)}(L-i),
$$
$\psi^{(v)}(\cdot)$ is the ordinary polygamma function expressed by
$$
\psi^{(v)}(L)=\frac{\partial^{v+1} \ln\Gamma(L)}{\partial L^{v+1}},
$$
for $v\geq 0$ (in this case, $\psi^{(0)}$ is known as the digamma function). 
By the linearity of the expectation operator, the following holds true:
$$
\operatorname{E}\big\{\operatorname{tr}(\boldsymbol{\Sigma}^{-1}\boldsymbol{Z})\big\}=\operatorname{tr}(\boldsymbol{\Sigma}^{-1}\operatorname{E}\{\boldsymbol{Z}\})=\operatorname{tr}\big(\boldsymbol{\Sigma}^{-1}\boldsymbol{\Sigma}\big)=m.
$$
Thus, the Shannon entropy relative to the random variable $\boldsymbol{Z}\sim \mathcal W_m(\boldsymbol{\Sigma},L)$ is expressed by
\begin{align}
H_\text{S}&(\boldsymbol{\theta})=\frac{m(m-1)}{2}\ln \pi- m^2 \ln L +m \ln |\boldsymbol{\Sigma}|+mL\nonumber \\
&\mbox{}+(m-L)\psi_m^{(0)}(L) +\displaystyle \sum_{k=0}^{m-1}\ln \Gamma(L-k).
\label{entropyw}
\end{align}
  
\subsubsection{Restricted Tsallis entropy}

Based on Table~\ref{entrpyexpression}, the restricted Tsallis entropy is defined by
$$
H_\text{T}^\beta(\boldsymbol{\theta})=(1-\beta)^{-1}(\widetilde{\mu}_\beta-1),
$$
where $\widetilde{\mu}_\beta\triangleq\operatorname{E}\bigl\{f_{\boldsymbol{Z}}^{\beta-1}(\boldsymbol{Z})\bigr\}$ can be explicitly calculated:
\begin{align}
\widetilde{\mu}_\beta&=\int_{\mathcal A}\Bigg[\frac{L^{mL}}{|\boldsymbol{\Sigma}|^L \Gamma_m(L)}\Bigg]^\beta |\boldsymbol{Z}'|^{\beta(L-m)}\\ \nonumber 
& \quad\quad \quad\quad \times\exp[-\beta L\operatorname{tr}(\boldsymbol{\Sigma}^{-1}\boldsymbol{Z}')] \mathrm{d}\boldsymbol{Z}'\nonumber\\
&=\Bigg[\frac{L^{mL}}{|\boldsymbol{\Sigma}|^L \Gamma_m(L)}\Bigg]^\beta \int_{\mathcal A}|\boldsymbol{Z}'|^{(\beta-1)(L-m)}|\boldsymbol{Z}'|^{L-m} \nonumber \\
&\quad\quad\quad\quad\quad\quad\quad\quad\times
\exp[-L\operatorname{tr}(\beta\boldsymbol{\Sigma}^{-1}\boldsymbol{Z}')] \mathrm{d}\boldsymbol{Z}'\nonumber \\
&=\Bigg[\frac{L^{mL}}{|\boldsymbol{\Sigma}|^L \Gamma_m(L)}\Bigg]^\beta\Bigg[\frac{|\boldsymbol{\Sigma}/\beta|^L \Gamma_m(L)}{L^{mL}}\Bigg]\operatorname{E}\big\{|\boldsymbol{X}|^{(1-\beta)(m-L)}\big\}\nonumber \\
&=\frac{L^{mL(\beta-1)}\beta^{-mL}}{|\boldsymbol{\Sigma}|^{L(\beta-1)}\Gamma_m^{\beta-1}(L)}\operatorname{E}\big\{|\boldsymbol{X}|^{(m-L)(1-\beta)}\big\},
\label{tsallis1}
\end{align}
where $\boldsymbol{X}\sim \mathcal W_m(\boldsymbol{\Sigma}/\beta,L)$.
Moreover, Anfinsen~\textit{et al.}~\cite{AnfinsenandEltoft2011} showed that 
\begin{align}
\operatorname{E}\bigl\{|\boldsymbol{Z}|^{s-m}\bigr\}=\frac{\Gamma_m(L+s-m)}{\Gamma_m(L)} \bigl(L^{-m}|\boldsymbol{\Sigma}|\bigr)^{s-m},
\label{tsallis2}
\end{align}
where $\boldsymbol{Z}\sim \mathcal W_m(\boldsymbol{\Sigma},L)$.
Thus, applying~\eqref{tsallis2} in \eqref{tsallis1}, we have that
\begin{equation}
\widetilde{\mu}_\beta=\frac{\Gamma_m(L+(1-\beta)(m-L))}{\Gamma^\beta_m(L)\beta^{m[L+(1-\beta)(m-L)]}}\frac{|\boldsymbol{\Sigma}|^{(1-\beta)m}}{L^{m^2(1-\beta)}}.
\label{exprforentro}
\end{equation}

\subsubsection{R\'enyi entropy}

From Table~\ref{entrpyexpression}, the R\'enyi entropy is given by 
$$
H_\text{R}^\beta(\boldsymbol{\theta})=(1-\beta)^{-1}\ln\widetilde{\mu}_{\beta}.
$$
Notice that this measure also depends on $\widetilde{\mu}_\beta$, which was already computed in Equation~\eqref{exprforentro}. 

Therefore, denoting $q=L+(1-\beta)(m-L)$, the R\'enyi entropy is expressed by
\begin{align}
&H_\text{R}^{\beta}(\boldsymbol{\theta})= \frac{m(m-1)}{2}\ln\pi- m^2 \ln L +m \ln |\boldsymbol{\Sigma}|\nonumber \\
&-\frac{mq\ln\beta}{1-\beta}+\frac{\sum_{i=0}^{m-1}\bigr[\ln\Gamma(q-i)-\beta\ln\Gamma(L-i)\bigl]}{1-\beta}.
\label{renyiw}
\end{align}

It is known that, as $\beta \rightarrow 1$, both the R\'enyi~\cite[p. 676]{coverandthomas1991} and Tsallis~\cite{Jiulin2010} entropies converge to the Shannon entropy. 
Thus:
\begin{equation}
\lim_{\beta \rightarrow 1}H_\text{R}^{\beta}(\boldsymbol{\theta}) = 
H_\text{S}(\boldsymbol{\theta})=
\lim_{\beta \rightarrow 1}H_\text{T}^{\beta}(\boldsymbol{\theta}).
\label{propetySR}
\end{equation}
These convergences hold true regardless the number of looks.
Moreover, it is important to emphasize that the derived expressions can be related to the eigenvalues of the covariance matrix.
This approach results in new expressions for $H_\text{S}$, $H_\text{T}^{\beta}$, and $H_\text{R}^{\beta}$ in terms of the geometrical and arithmetic mean of these eigenvalues as follows.
Let $\lambda_1,\lambda_2,\ldots,\lambda_p$ be the eigenvalues of the covariance matrix $\boldsymbol{\Sigma}$. 
Following Mardia et al~\cite{mardia}, $|\boldsymbol{\Sigma}|=\prod_{i=1}^m\lambda_i$ and $\operatorname{tr}(\boldsymbol{\Sigma})=\sum_{i=1}^m\lambda_i$, then 
$$
m\Big|\frac{\boldsymbol{\Sigma}}{\operatorname{tr}(\boldsymbol{\Sigma})}\Big|^{1/m}=\frac{\bigl(\prod_{i=1}^m\lambda_i\bigr)^{1/m}}{\bigl(\sum_{i=1}^m\lambda_i\bigr)/m}.
$$
Thus, setting $\boldsymbol{\Sigma}'={\boldsymbol{\Sigma}}/{\operatorname{tr}(\boldsymbol{\Sigma})}$ and adopting $\boldsymbol{\Sigma}'$ as the covariance matrix in~\eqref{entropyw} and~\eqref{renyiw}, we have new expressions that can be used in place of the ones proposed by Cloude and Pottier~\cite{CloudePottier1997} and Yan~\textit{et al.}~\cite{Yanetal2010}.

In the following, we examine the behavior of $H_\text{S}$, $H_\text{T}^{\beta}$, and $H_\text{R}^{\beta}$ in terms of $\boldsymbol{\Sigma}$ and~$L$.

\subsubsection{Case study}~\label{casestudy1}

\begin{figure*}[htb]
\centering
\subfigure[$H_\text{S}$ and $H^{\beta}_\text{R}$ for covariance matrix  $\boldsymbol{\Sigma}$ \label{entropyspeckle0}]{\includegraphics[width=.3\linewidth]{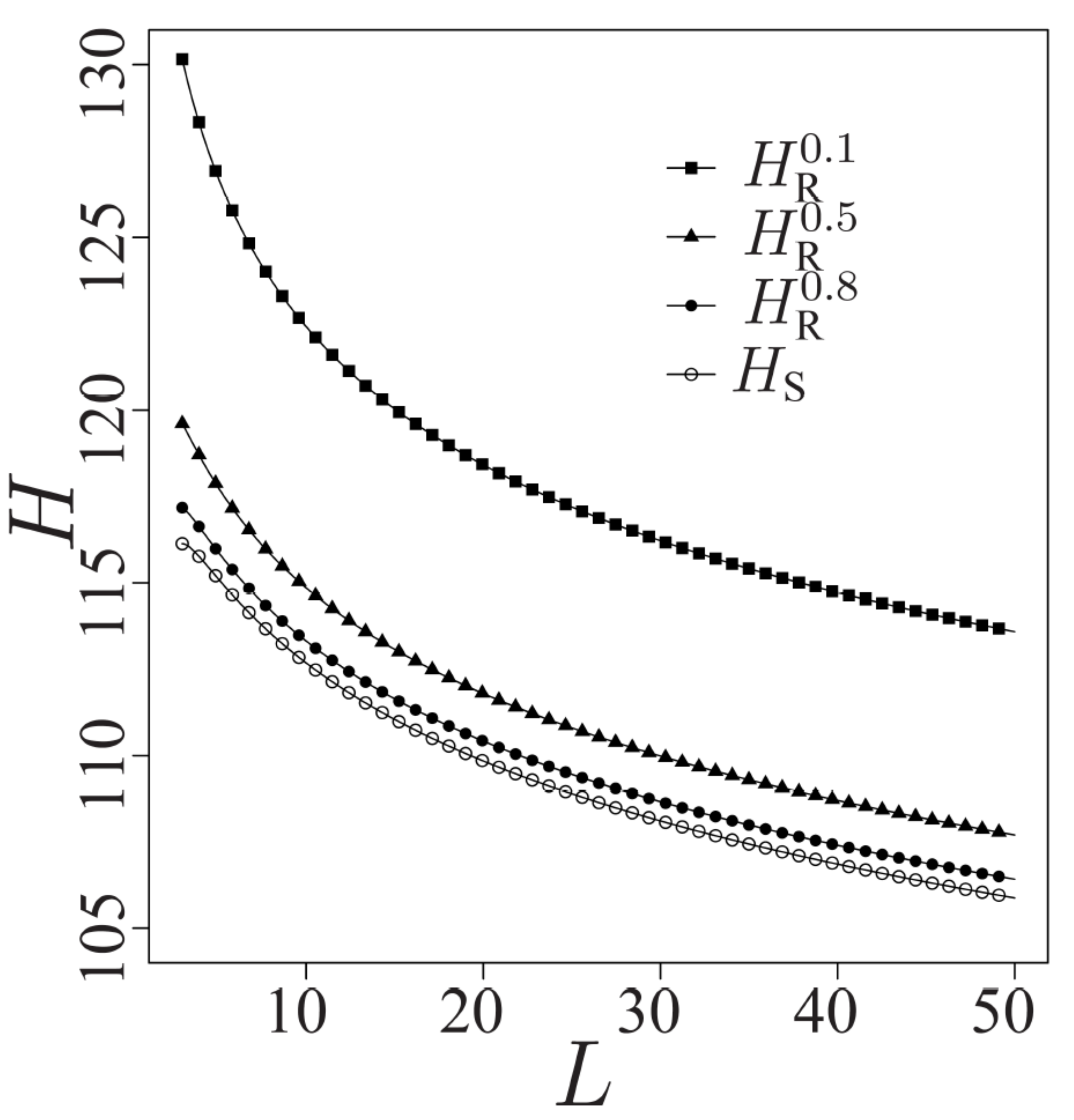}}
\subfigure[$H_\text{S}$ and $(1+k)\boldsymbol{\Sigma}$\label{entropyspeckle1}]{\includegraphics[width=.3\linewidth]{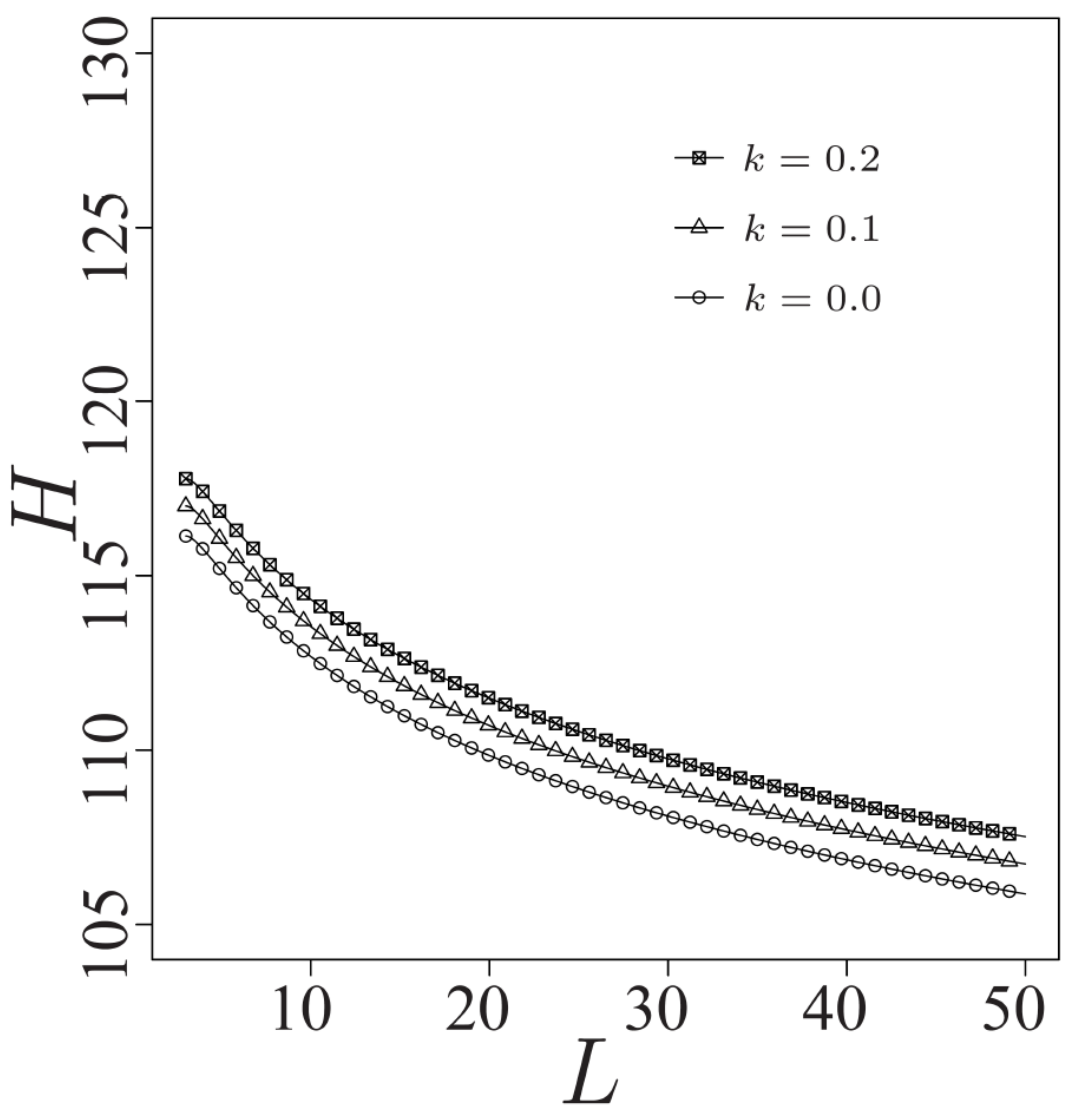}}
\subfigure[$H_\text{T}^{1-10^{-3}}$ and $(1+k)\boldsymbol{\Sigma}$\label{entropyspeckle4}]{\includegraphics[width=.3\linewidth]{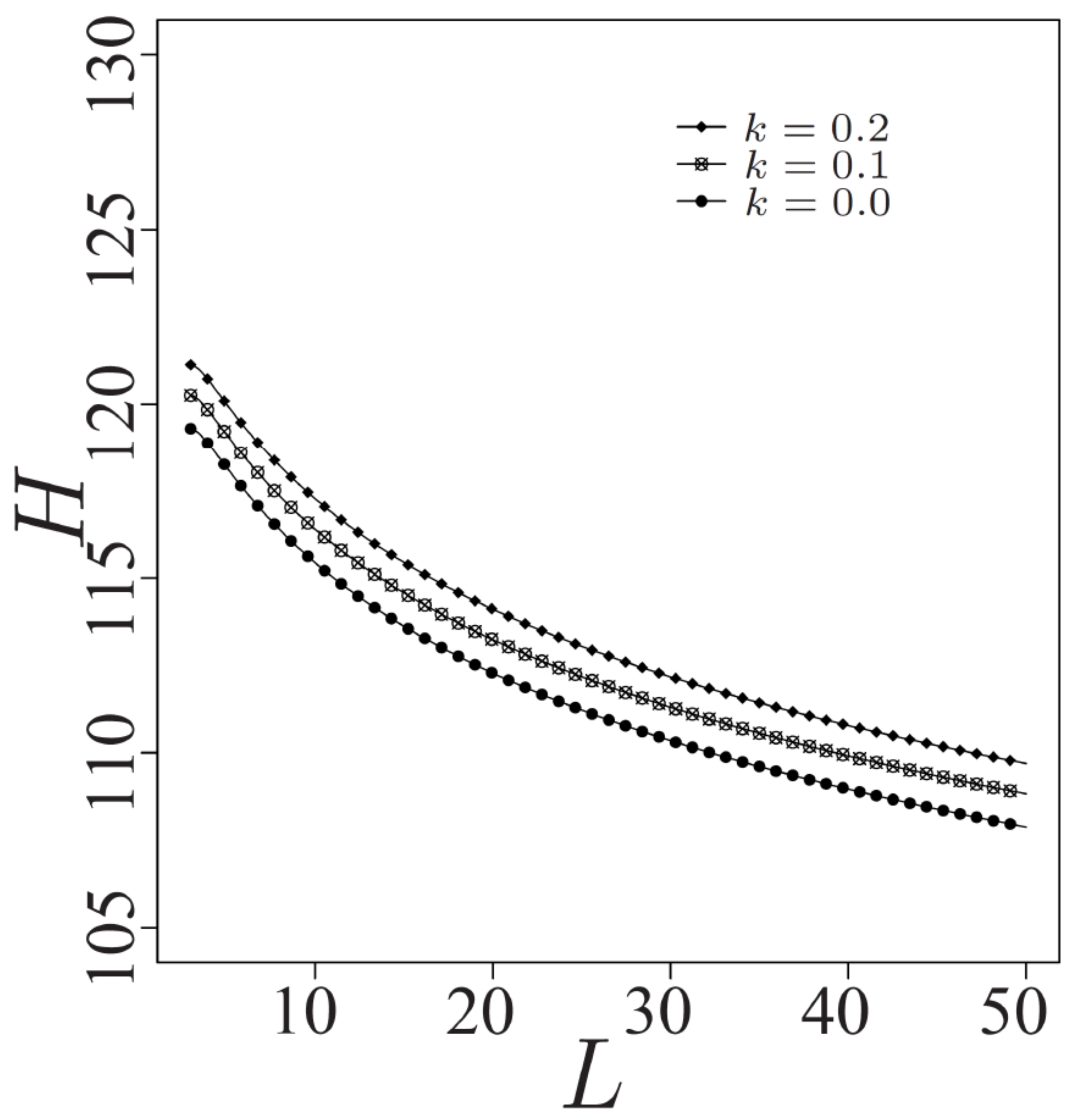}}
\subfigure[$H_\text{R}^{0.1}$ and $(1+k)\boldsymbol{\Sigma}$\label{entropyspeckle2}]{\includegraphics[width=.3\linewidth]{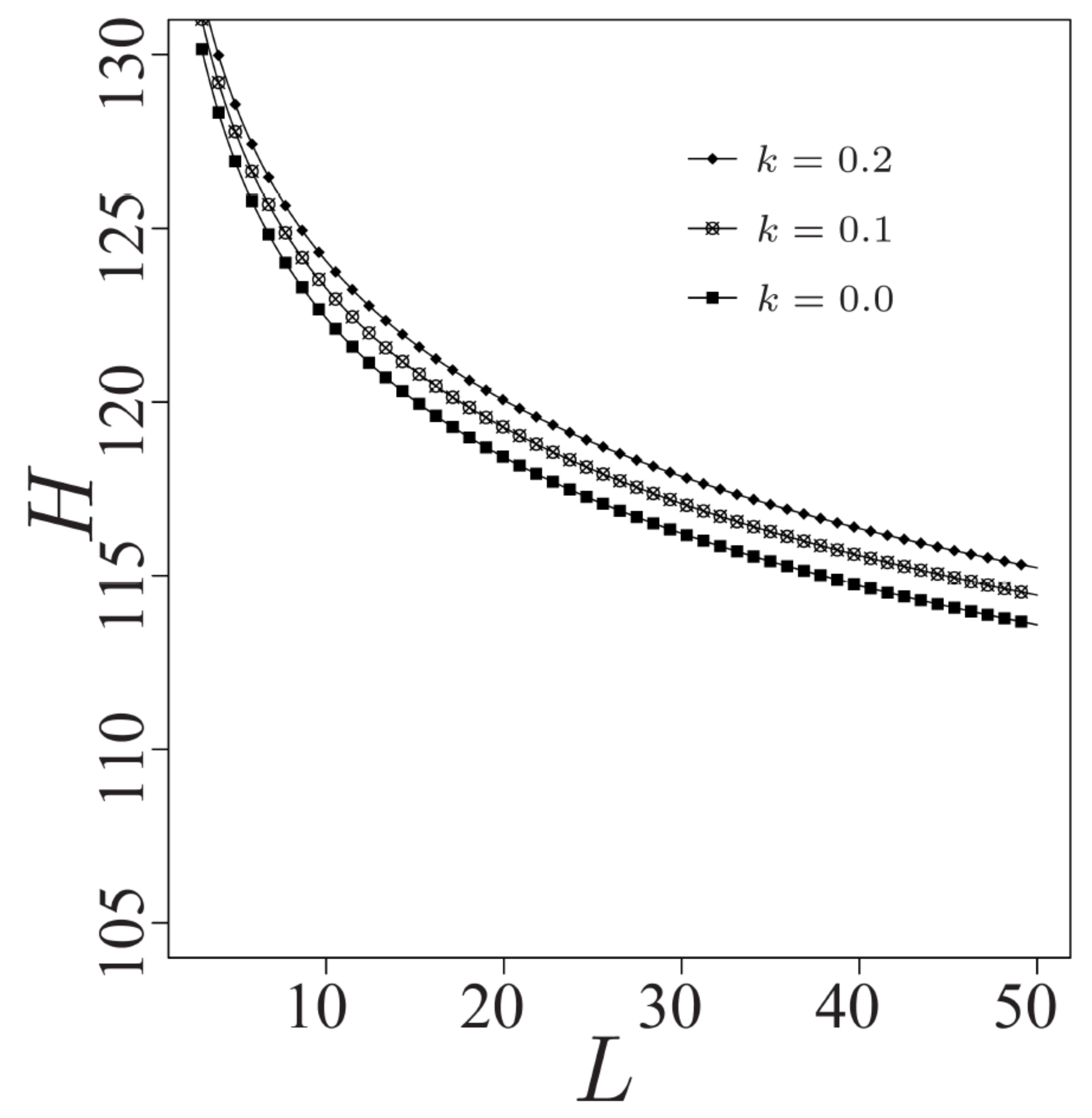}}
\subfigure[$H_\text{S}$ and $H_\text{T}^{\beta}$ and $\boldsymbol{\Sigma}$\label{entropyspeckle3}]{\includegraphics[width=.3\linewidth]{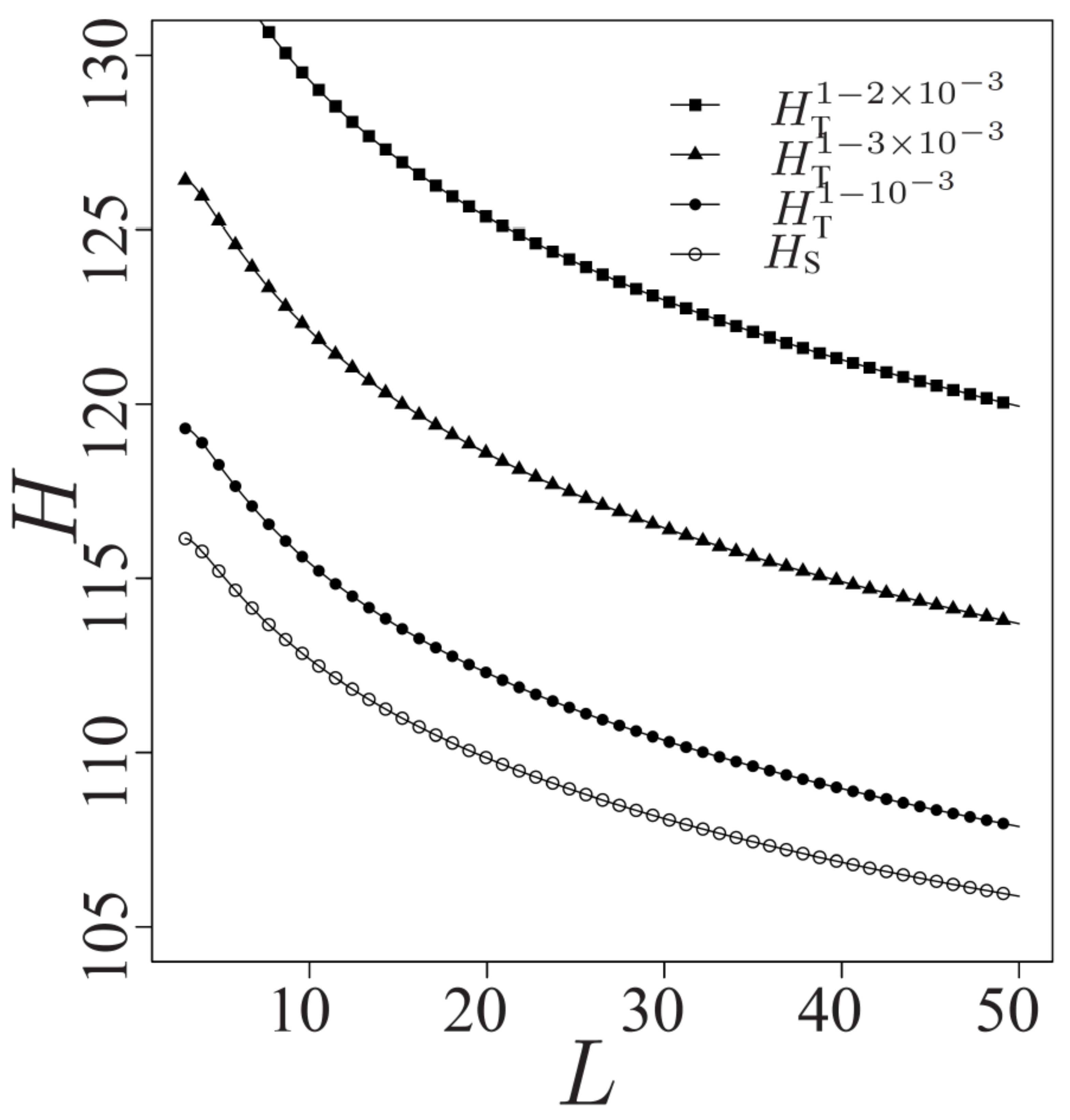}}
\label{renyifigure}
\caption{Shannon, Tsallis, and R\'enyi entropies for several covariance matrices and number of looks.}
\label{images3}  
\end{figure*}

Frery~\textit{et al.}~\cite{PolarimetricSegmentationBSplinesMSSP} observed the following covariance matrix on an urban area from the E-SAR image of We\ss ling (Bavaria, Germany):
$$ 
\boldsymbol{\Sigma}_U=
\left[ 
\begin{array}{ccc}
962892&19171-3579\textbf{i}&-154638+191388\textbf{i} \\
&56707&-5798+ 16812\textbf{i} \\
&&472251
\end{array} 
\right];
$$  
only the diagonal and the upper triangle values are shown.
Fig.~\ref{entropyspeckle0} depicts plots of the discussed entropies for $3\leq L \leq 50$, $\beta\in\{0.1,0.5,0.8\}$, and $\boldsymbol{\Sigma}_U$.
Considering the same interval for the number of looks, Figs.~\ref{entropyspeckle1},~\ref{entropyspeckle4}, and~\ref{entropyspeckle2} show the Shannon, Tsallis (of order $\beta=1-10^{-3}$), and R\'enyi (of order $\beta=0.1$) entropies for the covariance matrix $(1+k)\boldsymbol{\Sigma}_U$, $k\in\{0,0.1,0.2\}$, respectively.

Figs.~\ref{entropyspeckle0} and~\ref{entropyspeckle3} illustrate the property stated in~\eqref{propetySR}. 
In the case shown here, the convergences are from above, i.e., $H_\text{S}$ is always smaller than $H_\text{R}^\beta$ and $H_\text{T}^\beta$.

Figs.~\ref{entropyspeckle1} and~\ref{entropyspeckle2} suggest that multiplying the covariance matrix by a constant --- hence increasing its determinant --- also increases both Shannon and R\'enyi entropies.
As expected, increasing the number of looks leads to smaller entropy values due to the increased SNR.

Although the restricted Tsallis entropy can be used in several fields of image processing~\cite{MorioRefregierGoudailFernandezDupuis2009}, the derivation of its variance does not lead to a mathematically tractable expression.
Thus, henceforth we focus our attention on the Shannon and R\'enyi entropies, which allow the necessary algebraic manipulations for the the method described in Section~\ref{entropy:IT}.
In the next section, we derive asymptotic variances for the Shannon and R\'enyi entropy estimates.

\subsection{Asymptotic variances}\label{entropy:results2}

Let $\boldsymbol{Z}$ be a random matrix which follows a scaled complex Wishart distribution with parameter $\boldsymbol{\theta}$ as already defined.
Its log-likelihood is 
\begin{align*}
\ell(\boldsymbol{\theta})=&mL\ln L+(L-m)\ln |\boldsymbol{Z}|-L\ln |\boldsymbol{\Sigma}|\\
&-\frac{m(m-1)}{2}\ln \pi-\sum_{k=0}^{m-1}\ln \Gamma(L-k)-L\operatorname{tr}(\boldsymbol{\Sigma}^{-1}\boldsymbol{Z}).
\end{align*} 

Since $\partial \ell(\boldsymbol{\theta})/\partial \operatorname{vec}(\boldsymbol{\Sigma})=\operatorname{vec}(\partial \ell(\boldsymbol{\theta})/\partial \boldsymbol{\Sigma})$ holds true~\cite{HjorungnesandGesbert}, the score functions are given by
$$
\ell_L=m(\ln L+1)+\ln |\boldsymbol{Z}|-\ln |\boldsymbol{\Sigma}|
-\psi_m^{(0)}(L)-\operatorname{tr}\big(\boldsymbol{\Sigma}^{-1}\boldsymbol{Z}\big),
$$
and $\ell_{\operatorname{vec}(\Sigma)}=L \operatorname{vec}\big(\boldsymbol{\Sigma}^{-1}\boldsymbol{Z}\boldsymbol{\Sigma}^{-1}-\boldsymbol{\Sigma}^{-1}\big)$.

The Hessian matrix $\mathcal J(\boldsymbol{\theta})$, the Fisher information matrix $\mathcal K(\boldsymbol{\theta})$, and the biased version (according to Anfinsen \textit{et al.}~\cite{EstimationEquivalentNumberLooksSAR}) for Cram\'er-Rao lower bound $\mathcal C(\boldsymbol{\theta})$ are necessary to obtain closed form expressions for the asymptotic entropy variance used in Equation~\eqref{varentro}.
In particular, the following quantity plays a central role:
$$
\mathcal J_{\Sigma\Sigma}=\frac{\partial}{\partial \operatorname{vec}(\boldsymbol{\Sigma})^*}\operatorname{vec}\bigg(\frac{\partial \ell(\boldsymbol{\theta})}{\partial \boldsymbol{\Sigma}}\bigg)^t,
$$
where $(\cdot)^*$ represents complex conjugation.	
Anfinsen~\textit{et al.}~\cite{EstimationEquivalentNumberLooksSAR} showed that 
\begin{align*}
T_1&=-\frac{\partial (\boldsymbol{\Sigma}^{-1}\boldsymbol{Z}\boldsymbol{\Sigma}^{-1})}{\partial \boldsymbol{\Sigma}}\\
&=-\boldsymbol{\Sigma}^{-1} \otimes \boldsymbol{\Sigma}^{-1}\boldsymbol{Z}\boldsymbol{\Sigma}^{-1}-\boldsymbol{\Sigma}^{-1} \otimes \boldsymbol{\Sigma}^{-1}\boldsymbol{\Sigma}^{-1}\boldsymbol{Z}.
\end{align*}
Moreover, it is known that~\cite{HjorungnesandGesbert} 
$$T_2=\partial \boldsymbol{\Sigma}^{-1}/\partial \boldsymbol{\Sigma}= - \boldsymbol{\Sigma}^{-1} \otimes \boldsymbol{\Sigma}^{-1}.$$
Thus, we have that
\begin{align}\label{hessian}
\mathcal J_{\Sigma\Sigma}=&L(T_2-T_1)=L(\boldsymbol{\Sigma}^{-1} \otimes \boldsymbol{\Sigma}^{-1}
-\boldsymbol{\Sigma}^{-1} \otimes \boldsymbol{\Sigma}^{-1}\boldsymbol{Z}\boldsymbol{\Sigma}^{-1} \nonumber \\
&-\boldsymbol{\Sigma}^{-1} \otimes \boldsymbol{\Sigma}^{-1}\boldsymbol{\Sigma}^{-1}\boldsymbol{Z}).
\end{align}
From Equation~\eqref{hessian} we obtain:
\begin{align}\label{fisherinformation}
\mathcal K_{\Sigma\Sigma}=\operatorname{E}\{-\mathcal J_{\Sigma\Sigma}\}=L \boldsymbol{\Sigma}^{-1} \otimes \boldsymbol{\Sigma}^{-1}.
\end{align}

The analytical expressions for $\mathcal J(\boldsymbol{\theta})$,  $\mathcal K(\boldsymbol{\theta})$, and $\mathcal C(\boldsymbol{\theta})$ are, thus,
\begin{align*}
&\mathcal J(\boldsymbol{\theta})=\\
&\quad\Bigg[ \begin{array}{cc}
\frac{m}{L} - \psi_m^{(1)}(L)& \operatorname{vec}(\boldsymbol{\Sigma}^{-1}\boldsymbol{Z}\boldsymbol{\Sigma}^{-1}-\boldsymbol{\Sigma}^{-1})^t \\
\operatorname{vec}(\boldsymbol{\Sigma}^{-1}\boldsymbol{Z}\boldsymbol{\Sigma}^{-1}-\boldsymbol{\Sigma}^{-1})^* & \mathcal J_{\Sigma \Sigma} 
\end{array} \Bigg], \\
&\mathcal K(\boldsymbol{\theta})=\operatorname{E}\{-\mathcal J(\boldsymbol{\theta})\}= \\
&\quad\bigg[ \begin{array}{cc}
 \psi_m^{(1)}(L) - \frac{m}{L}& \operatorname{vec}(\boldsymbol{0}_m)^t \\
 \operatorname{vec}(\boldsymbol{0}_m) & L \boldsymbol{\Sigma}^{-1} \otimes \boldsymbol{\Sigma}^{-1} 
\end{array} \bigg],\\
&\mathcal C(\boldsymbol{\theta})=\mathcal K(\boldsymbol{\theta})^{-1}=\\
&\quad\left[ \begin{array}{cc}
 [\psi_m^{(1)}(L) - \frac{m}{L}]^{-1} & \operatorname{vec}(\boldsymbol{0}_m)^t \\
 \operatorname{vec}(\boldsymbol{0}_m)  & L^{-1}\boldsymbol{\Sigma} \otimes \boldsymbol{\Sigma}
\end{array} \right],
\end{align*}
where $\boldsymbol{0}_{m}$ is the null square matrix of order $m$. 

Anfinsen~\textit{el al.}~\cite{EstimationEquivalentNumberLooksSAR} derived the Fisher information matrix for the unscaled complex Wishart law.
That approach found that the parameters of such distribution are not orthogonal.
Based on $\mathcal J(\boldsymbol{\theta})$, we conclude that  $L$ and $\boldsymbol{\Sigma}$ become orthogonal under the scaling of the complex Wishart law.
Among other benefits, such scaling makes the likelihood equations separable, as shown in the following.

Consider $\{\boldsymbol{Z}_1,\boldsymbol{Z}_2,\ldots,\boldsymbol{Z}_N\}$ a random sample of size $N$ obtained from $\boldsymbol{Z}\sim \mathcal W_m(\boldsymbol{\Sigma},L) $. 
Since $N^{-1}\sum_{k=1}^N \nabla \ell_k(\widehat{\boldsymbol{\theta}})=\boldsymbol{0}$, we have that
$$
\widehat{\boldsymbol{\Sigma}}=\frac{1}{N}\sum_{k=1}^N {\boldsymbol{Z}_k}=\overline{\boldsymbol{Z}},
$$
and
\begin{align}
m\ln\widehat{L}+\frac 1N\sum_{k=1}^N \ln|\boldsymbol{Z}_k|-\ln|\overline{\boldsymbol{Z}}|-\psi_m^{(0)}(\widehat{L})=0.
\label{eqscore1}
\end{align}
Thus, the ML estimator of $\boldsymbol{\Sigma}$ is the sample mean, while $\widehat L$ is obtained solving the system shown in Equation~\eqref{eqscore1}. 
The Newton-Raphson iterative method~\cite{gentle2002elements} can be used to solve this nonlinear system.

The asymptotic variance given by Equation~\eqref{varentro} is determined by $\mathcal C(\boldsymbol{\theta})$ and by the term~\cite{Pardo1997} 
$$
\boldsymbol{\delta}=\biggl[\frac{\partial H^h_{\phi}(\boldsymbol{\theta})}{\partial L}\;\operatorname{vec}\biggl(\frac{H^h_{\phi}(\boldsymbol{\theta})}{\partial \boldsymbol{\Sigma}}\biggr)^t\biggr]^t.
$$
We denote the resulting $\boldsymbol{\delta}$ as $\boldsymbol{\delta}_\text{S}$ and $\boldsymbol{\delta}_{\text{R},\beta}$, when $H_\text{S}$ and $H^{\beta}_\text{R}$ are considered, respectively.
We could not find a closed expression for the variance of the restricted Tsallis entropy, so it will not be further considered in the remainder of this work.
Analogously, entropy variances are denoted as $\sigma_{\text{S}}^2$ and $\sigma_{\text{R},\beta}^2$.
These quantities are given by expressions \eqref{exvarianceentropyshannon}-\eqref{varianceentropyrenyi}.

\begin{figure*}
\begin{itemize}
\item [(a)] Shannon entropy:
{
\begin{equation}
\boldsymbol{\delta}_\text{S}=\Bigg[ \begin{array}{c}
(m-L)\psi_m^{(1)}(L)+m-\frac{m^2}{L}\\
m \operatorname{vec}\bigl(\boldsymbol{\Sigma}^{-1}\bigr)
\end{array} \Bigg]
\label{exvarianceentropyshannon}
\end{equation}
}
and
{
\begin{align}
\sigma_{\text{S}}^2=\frac{\bigl[(m-L)\psi_m^{(1)}(L)+m-\frac{m^2}{L}\bigr]^2}{\psi_m^{(1)}(L) - \frac{m}{L}}
+\frac{m^2}{L} \operatorname{vec}\bigl(\boldsymbol{\Sigma}^{-1}\bigr)^t \bigl( \boldsymbol{\Sigma} \otimes \boldsymbol{\Sigma}\bigr)  \operatorname{vec}\bigl(\boldsymbol{\Sigma}^{-1}\bigr).
\label{varianceentropyshannon}
\end{align}
}
\item [(b)] R\'enyi entropy:
{
\begin{equation}
\boldsymbol{\delta}_{\text{R},\beta}=\Bigg[ \begin{array}{c}
\frac{\beta}{1-\beta} \bigl[\psi_m^{(0)}(q)-\psi_m^{(0)}(L)\bigr]-\frac{m\beta\ln(\beta)}{1-\beta}-\frac{m^2}{L}\\
m \operatorname{vec}\bigl(\boldsymbol{\Sigma}^{-1}\bigr)
\end{array} \Bigg]
\label{exvarianceentropyrenyi}
\end{equation}
}
and
{
\begin{align}
\sigma_{\text{R},\beta}^2=
\frac{\Big\{\frac{\beta}{1-\beta} \bigl[\psi_m^{(0)}(q)-\psi_m^{(0)}(L)\bigr]-\frac{m\beta\ln(\beta)}{1-\beta}-\frac{m^2}{L}\Big\}^2}{\psi_m^{(1)}(L) - \frac{m}{L}}
+\frac{m^2}{L} \operatorname{vec}\bigl(\boldsymbol{\Sigma}^{-1}\bigr)^t \bigl(\boldsymbol{\Sigma} \otimes \boldsymbol{\Sigma}\bigr)\operatorname{vec}\bigl(\boldsymbol{\Sigma}^{-1}\bigr). 
\label{varianceentropyrenyi}
\end{align}
}
\end{itemize}
\hrulefill
\end{figure*}

Note that, as expected from Equation~\eqref{propetySR}, 
$
\lim_{\beta \rightarrow 1}\sigma_{\text{R},\beta}^2 = \sigma_{\text{S}}^2.
$

The entropies, along with their variances, can be used as alternative goodness-of-fit tests to the one proposed in~\cite{AnfinsenDoulgerisEltoft2011} for the Wishart distribution, specifying $r=1$ in Equation~\eqref{cocran0}.
Cintra \textit{et al.}~\cite{ParametricNonparametricTestsSpeckledImagery} showed that these statistics outperform the Kolmogorov-Smirnov nonparametric test in intensity SAR imagery. 
In this paper, we aim to quantify contrast in PolSAR images and situations with $r=2$.
In the following, the proposed tests are applied to PolSAR imagery.

\section{Applications}

In this section, we combine the entropies and variances derived in the previous section to form statistical tests, whose performance with respect to test size and power is assessed by Monte Carlo experiments.
Finally, the discussed methodology is applied to real data.

\subsection{Entropy as a feature for discrimination}

\begin{figure}[htb]
\centering
\subfigure[E-SAR intensity image (HH band) \label{ImageAP11}]{\includegraphics[width=.5\linewidth]{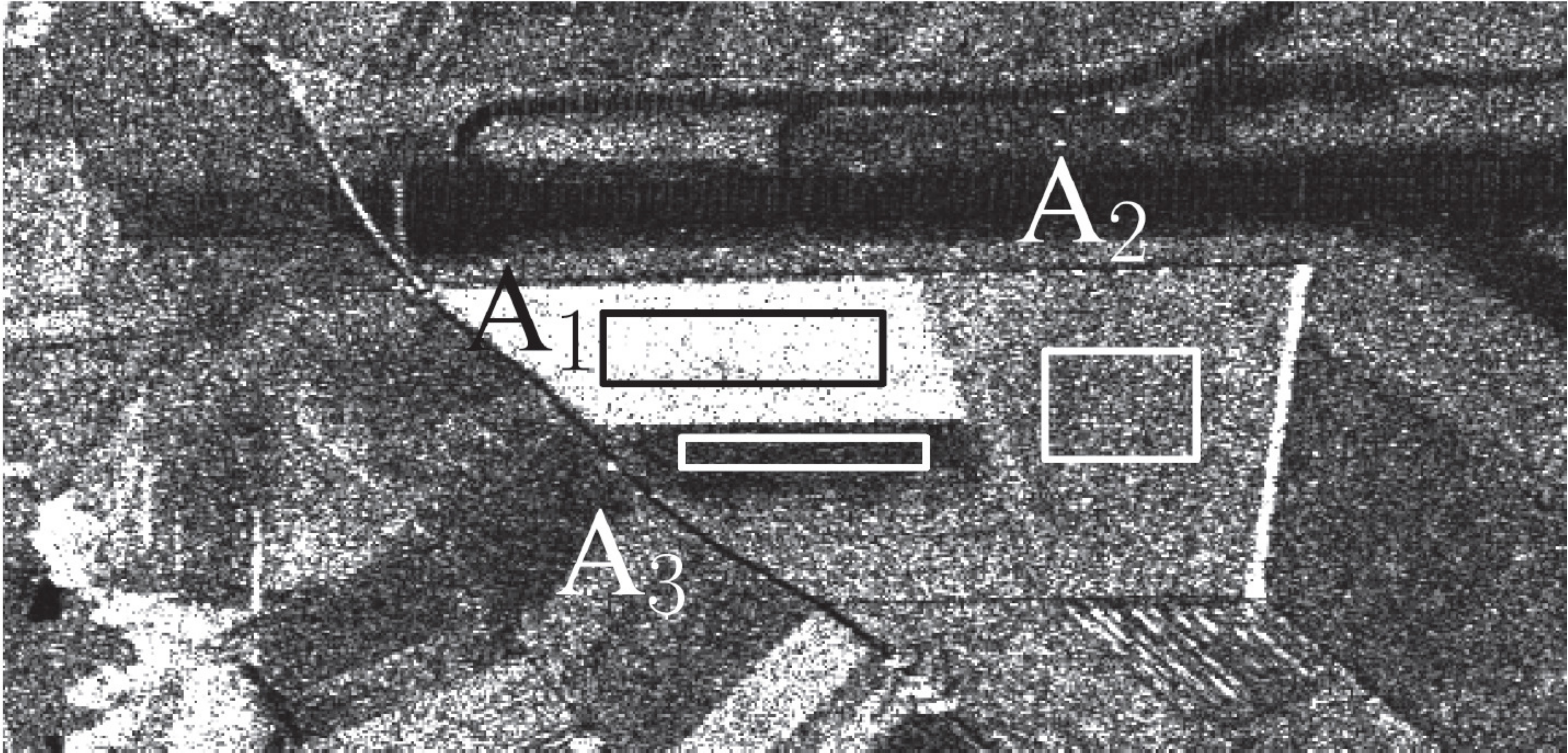}}
\subfigure[Data from region $\text{A}_1$\label{ImageAP12}]{\includegraphics[width=.48\linewidth]{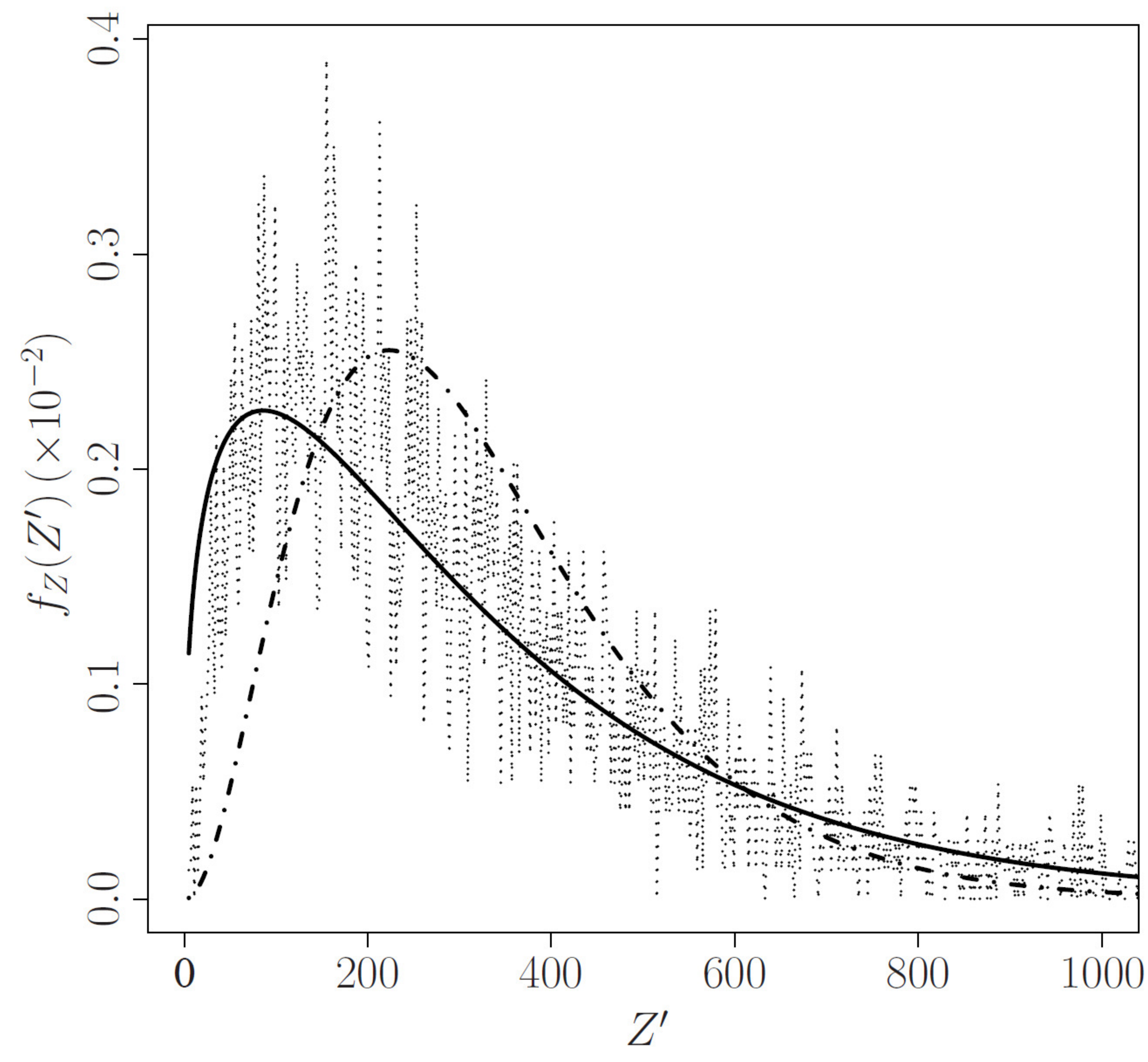}}
\subfigure[Data from region $\text{A}_2$\label{ImageAP13}]{\includegraphics[width=.48\linewidth]{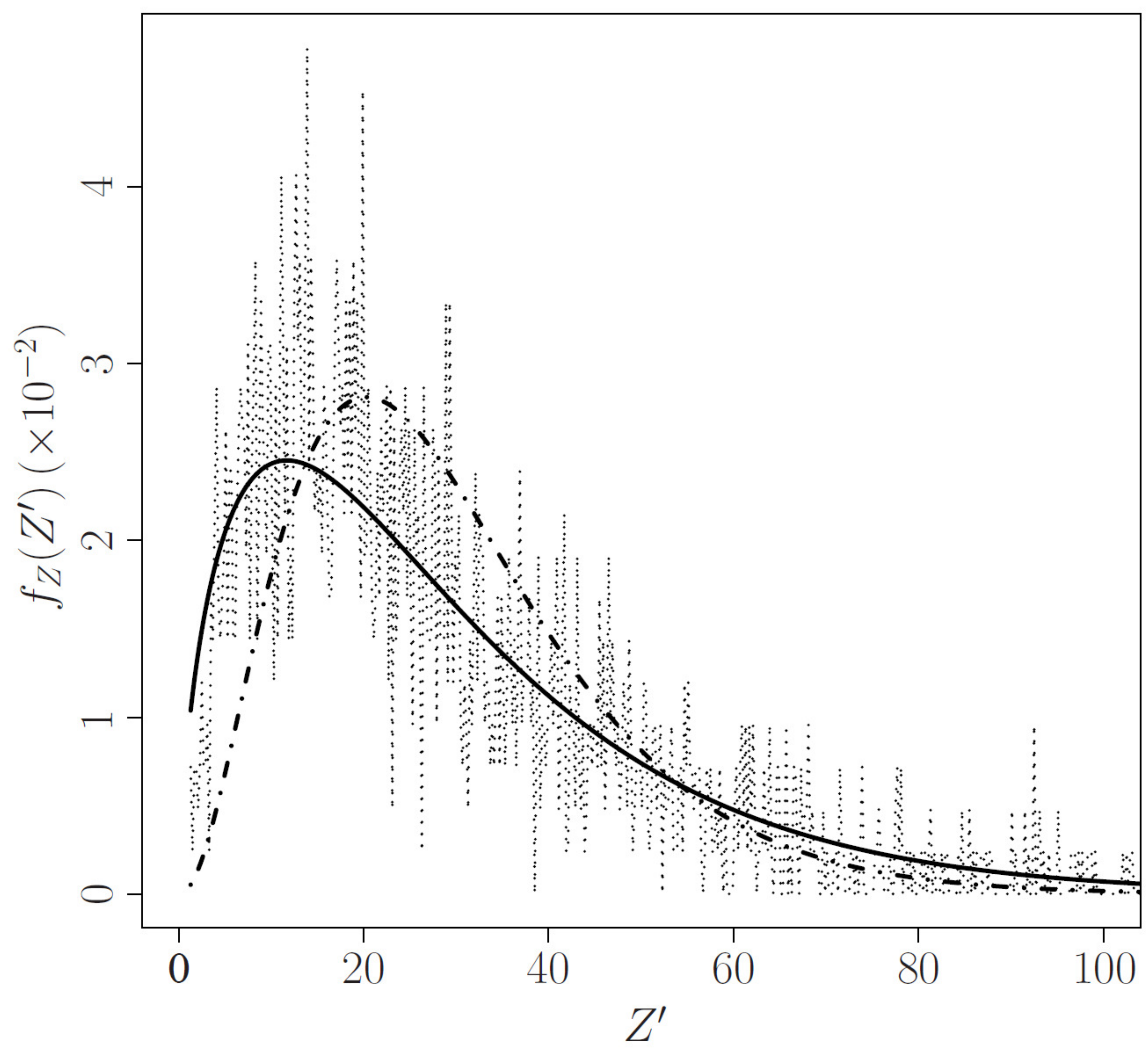}}
\subfigure[Data from region $\text{A}_3$ \label{ImageAP14}]{\includegraphics[width=.48\linewidth]{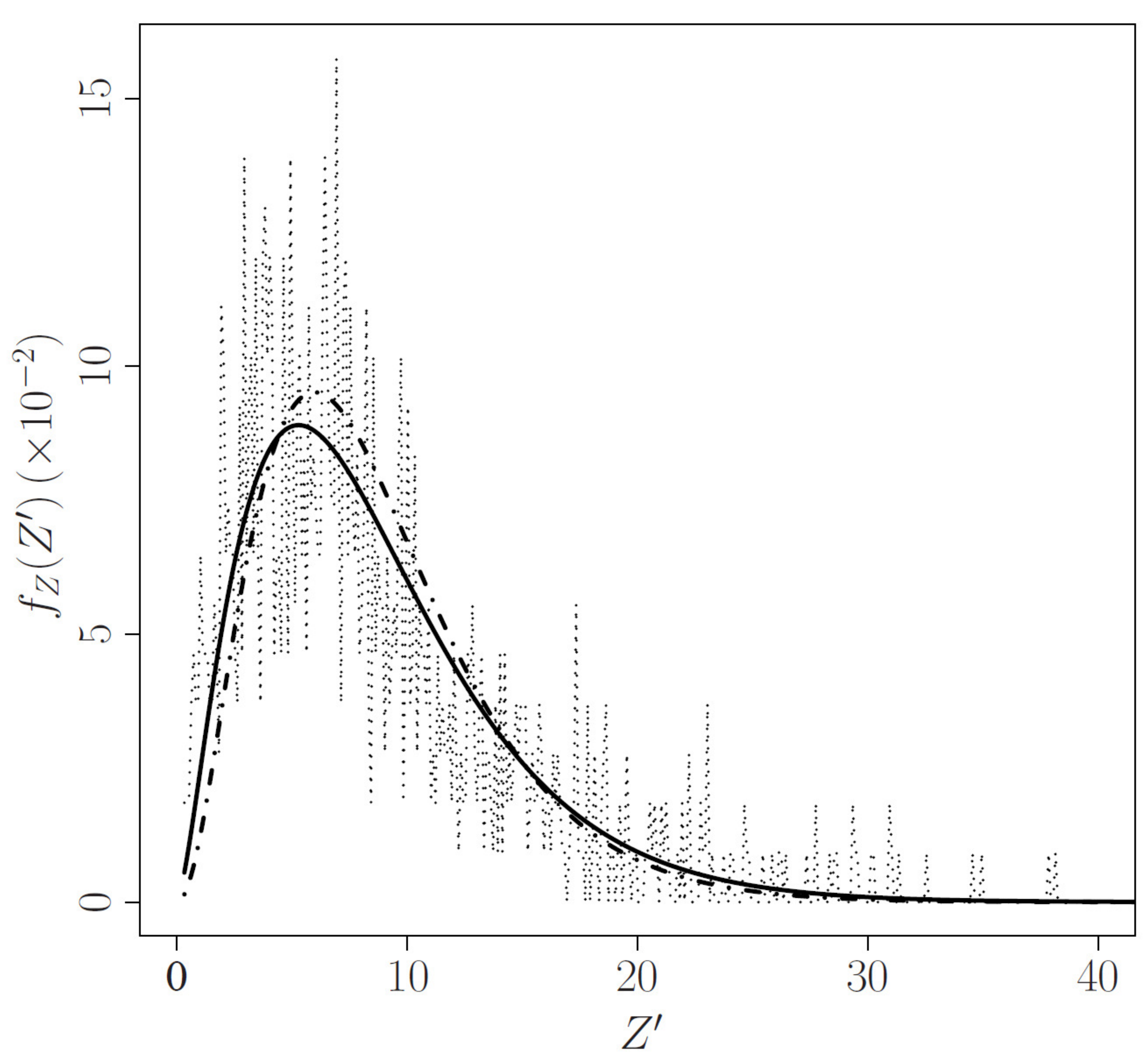}}
\caption{PolSAR image, selected regions, histograms (dot curve) and fitted scaled complex Wishart distributions: ${\mathcal W_R}(\widehat{\boldsymbol{\Sigma}},\widehat{L})$ (solid curve) and ${\mathcal W}(\widehat{\boldsymbol{\Sigma}},3.2)$ (dashed curve).} 
\label{ImageAP1}  
\end{figure}

Fig.~\ref{ImageAP11} shows the HH band of a PolSAR L-band image over We\ss ling, Germany.
This image was obtained by the E-SAR sensor~\cite{ESAR} with $3.2$ equivalent number of  looks (ENL), which is a PolSAR parameter associated to the degree of averaging of SAR measurements during data formation~\cite{EstimationEquivalentNumberLooksSAR}.
Three regions were selected: $\text{A}_1$, $\text{A}_2$, and $\text{A}_3$, corresponding to areas with strong, moderate, and weak return, respectively.
The image has approximately \unit[3]{m} resolution.

Table~\ref{tableapplication1} shows the sample size $N$ and the ML estimates of the scaled Wishart distribution parameters for each region.
The ENL estimates are close, suggesting similar levels of radar texture in the three regions; such levels are probably low, since the regions appear to be cropland.
Notice that the smallest estimated number of looks and the highest value of determinant for the sample convariance matrix belong to region $\text{A}_1$, where the texture is more pronounced.  
Figs.~\ref{ImageAP12}-\ref{ImageAP14} present histograms and fitted densities from considered data for two distributions of the scaled complex Wishart family: ${\mathcal W}(\widehat{\boldsymbol{\Sigma}},3.2)$ and ${\mathcal W_R}(\widehat{\boldsymbol{\Sigma}},\widehat{L})$.
Additionally, Table~\ref{tableapplication1} also presents the Akaike information criterion (AIC), a measure of the goodness of these fits~\cite{SeghouaneAmari2007}.
In all cases, the ${\mathcal W_R}$ distribution presented the best fit.

\begin{table}[hbt]                                                                                      
\centering            
\caption{ML estimates for the samples from Figure~\protect{\normalfont{\ref{ImageAP11}}} under the relaxed and original Wishart distributions, and their AIC values
}\label{tableapplication1}
\begin{tabular}{rrrrrr}
\toprule                                                           
\multirow{2}{*}{Regions} & \multirow{2}{*}{$N$}   & \multirow{2}{*}{$|\widehat{\boldsymbol{\Sigma}}|$} &  \multirow{2}{*}{$\widehat{L}$} & \multicolumn{2}{c}{AIC}
\\  \cmidrule(lr{.25em}){5-6}
&&&& ${\mathcal W}(\widehat{\boldsymbol{\Sigma}},3.2)$& ${\mathcal W_R}(\widehat{\boldsymbol{\Sigma}},\widehat{L}) $\\
\midrule
$\text{A}_1$ &  3708 &  355494.500 & 1.361 & 50769.93 & 49856.90\\
$\text{A}_2$ &  2088 &  3321.241 & 1.657 & 18353.35 & 17931.51\\
$\text{A}_3$ &  1079  & 274.189 & 2.557 & 6749.56 & 6629.15\\
\bottomrule
\end{tabular}
\end{table}

Table~\ref{tableapplication22} presents the asymptotic lower and upper bounds for the Shannon and R\'enyi entropies at the 95\% level of confidence, as computed according to Equation~\eqref{intervalforone}. 
The intervals are disjoint, suggesting that the entropy can be used as a feature for region discrimination.
The smallest entropies are associated with regions whose estimated covariance matrices have small determinants, in accordance with the case study presented in Section~\ref{casestudy1}.

Noting that (i)~the entropy is a measure of randomness, which is associated to variability, and that (ii)~PolSAR areas with high reflectance are more affected by speckle noise, even those with negligible texture, it is intuitive that the determinant of the covariance matrix and the entropy are positively correlated.
Moreover, the expressions of entropies are also directly proportional to the determinant, i.e., $H_{\text{R}}^{\beta},H_{\text{S}}\propto m\ln|\boldsymbol{\Sigma}|$.

Goodman~\cite{Goodmana} defined the stochastic covariance matrix determinant as a \textit{generalized variance} which is associated with the speckle variability, defined as the effect of the speckle noise resulting from multipath interference.
Additionally, when there is texture variability it is due to the spatial variability in reflectance, and is associated with the ``heterogeneity'', which is the usual measure of the texture level.
This last source of variability can be captured by, for instance, the roughness parameter of the polarimetric $\mathcal G^0$ law~\cite{FreitasFreryCorreia:Environmetrics:03,FreryCorreiaFreitas:ClassifMultifrequency:IEEE:2007}.

\begin{table}[hbt]                                                                                      
\centering                                                                                                                                                                                         
\caption{Estimated asymptotic intervals at 95\%  for the Shannon and R\'enyi entropies of samples from Figure~\protect{\normalfont{\ref{ImageAP11}}}}\label{tableapplication22}
\begin{tabular}{rrrrrrr}
\toprule                                                           
\multirow{2}{*}{\rotatebox{90}{Region}} & \multicolumn{2}{c}{$\widehat{H_\text{S}}$} & \multicolumn{2}{c}{$\widehat{H_\text{R}^{0.1}}$} & \multicolumn{2}{c}{$\widehat{H_\text{R}^{0.8}}$}\\
 \cmidrule(lr{.25em}){2-3}\cmidrule(lr{.25em}){4-5}\cmidrule(lr{.25em}){6-7}
 & Lower  & Upper & Lower & Upper & Lower  & Upper \\
\cmidrule(lr{.25em}){2-3}\cmidrule(lr{.25em}){4-5}\cmidrule(lr{.25em}){6-7}
$\text{A}_1$ & 37.979 & 38.432 & 61.083 & 61.332 & 44.045 & 44.364 \\
$\text{A}_2$ & 30.079 & 30.541 & 45.563 & 45.867 & 36.124 & 37.049 \\
$\text{A}_3$ & 19.611 & 19.949 & 35.000 & 35.346 & 20.901 & 21.230 \\
\bottomrule
\end{tabular}
\end{table}

\subsection{Synthetic Data}

We quantify the performance of hypothesis tests based on Shannon and R\'enyi entropies with synthetic data following the scaled complex Wishart distribution, generated as suggested in~\cite{leeetal1994a}. 

The simulation employed the following parameters: (i)~$L=3.2$; (ii)~the estimated covariance matrices from regions $\text{A}_1$, $\text{A}_2$, and $\text{A}_3$, denoted as $\widetilde{\boldsymbol{\Sigma}}_1, \widetilde{\boldsymbol{\Sigma}}_2$, and $\widetilde{\boldsymbol{\Sigma}}_3$, respectively; (iii)~sample sizes $N=\{9,49,81,121,400\}$ (i.e., squared windows of side $\{3, 7,9, 11, 20\}$ pixels); and (iv)~significance levels $\alpha\in \{1\%,5\%,10\%\}$.
After generating two samples of size $N$ from $\boldsymbol{X}\sim \mathcal W_m(\widetilde{\boldsymbol{\Sigma}}_i,L)$ and $\boldsymbol{Y}\sim \mathcal W_m(\widetilde{\boldsymbol{\Sigma}}_j,L)$, we tested the the null hypothesis $\mathcal H_0 \colon (\widetilde{\boldsymbol{\Sigma}}_i,L)=(\widetilde{\boldsymbol{\Sigma}}_j,L)$ for $i,j=1,2,3$.
Following~\cite{HypothesisTestingSpeckledDataStochasticDistances}, we run $5500$ replicas of the Monte Carlo simulation for every case.
Empirical test size and power were employed as figures of merit.
These quantities are defined as the rejection rates of $\mathcal H_0$ when this hypothesis is true and is false (i.e., $\boldsymbol{\Sigma}_i \neq \boldsymbol{\Sigma}_j$), respectively; they are also called Type~I error and true positive rates.

Table~\ref{tableapplication2} presents the resulting empirical test sizes.
From a purely statistical viewpoint, the ideal test is the one whose empirical size is exactly the nominal one for simulated data.
Therefore, we conclude that the R\'enyi entropy for $\beta=0.1$ is generally outperformed by the Shannon and R\'enyi of order $0.8$, except for $N=9$.
The two last hypothesis tests have empirical test sizes close to the nominal levels even for small sample sizes.

\begin{table}[htb]
\centering
\caption{Empirical test sizes based on Shannon and R\'enyi (of order $0.1$ and $0.8$) entropies for regions A$_1$, A$_2$, and A$_3$ from Figure~\protect{\normalfont{\ref{ImageAP11}}}; a good test is close to its nominal level.}\label{tableapplication2}
\begin{tabular}{r@{ } r@{\quad}r@{\quad}r@{\quad} r@{\quad}r@{\quad}r@{\quad} r@{\quad}r@{\quad}r@{\quad}}
\toprule
& \multicolumn{3}{c}{$\text{A}_1$} & \multicolumn{3}{c}{$\text{A}_2$} & \multicolumn{3}{c}{$\text{A}_3$} \\ 
\cmidrule(lr{.25em}){2-4} \cmidrule(lr{.25em}){5-7} \cmidrule(lr{.25em}){8-10}
$N$ & $1\%$ & $5\%$ & $10\%$  & $1\%$ & $5\%$ & $10\%$ & $1\%$&$5\%$ & $10\%$ \\ 
\cmidrule(lr{.25em}){1-1} \cmidrule(lr{.25em}){2-4} \cmidrule(lr{.25em}){5-7} \cmidrule(lr{.25em}){8-10}
& \multicolumn{9}{c}{Shannon entropy}\\
9   & 1.35	&	6.15	&	11.76	&	1.36	&	6.16	&	12.02	&	1.44	&	6.20	&	12.05  \\ 
49  & 1.38	&	5.53	&	10.44	&	1.38	&	5.65	&	10.35	&	1.35	&	5.47	&	10.44  \\ 
81  & 1.33	&	5.51	&	9.75	&	1.35	&	5.47	&	9.84	&	1.35	&	5.36	&	10.07  \\ 
121 & 1.18	&	5.56	&	10.58	&	1.15	&	5.51	&	10.51	&	1.15	&	5.49	&	10.44  \\ 
400 & 1.35	&	5.38	&	10.33	&	1.38	&	5.47	&	10.40	&	1.40	&	5.51	&	10.22  \\ 
\cmidrule(lr{.25em}){1-1} \cmidrule(lr{.25em}){2-10}
& \multicolumn{9}{c}{R\'enyi entropy with $\beta=0.8$ }\\
9 & 1.38	&	6.22	&	11.65	&	1.44	&	6.18	&	11.93	&	1.45	&	6.09	&	11.78  \\     
49 & 1.27	&	5.62	&	10.36	&	1.31	&	5.65	&	10.31	&	1.33	&	5.38	&	10.55  \\     
81 & 1.31	&	5.47	&	9.76	&	1.20	&	5.47	&	9.91	&	1.29	&	5.36	&	9.95   \\     
121 & 1.05	&	5.35	&	10.45	&	1.15	&	5.44	&	10.44	&	1.13	&	5.53	&	10.55  \\     
400 & 1.35	&	5.27	&	10.31	&	1.36	&	5.29	&	10.29	&	1.27	&	5.33	&	10.42  \\     
\cmidrule(lr{.25em}){1-1} \cmidrule(lr{.25em}){2-10}
& \multicolumn{9}{c}{R\'enyi entropy with $\beta=0.1$ }\\
9 & 0.69	&	4.04	&	8.80	&	0.76	&	4.15	&	8.65	&	0.82	&	4.04	&	8.55   \\     
49 & 0.71	&	3.62	&	7.82	&	0.75	&	3.87	&	7.89	&	0.69	&	3.45	&	7.73   \\     
81 & 0.53	&	3.69	&	7.51	&	0.56	&	3.84	&	7.49	&	0.55	&	3.62	&	7.51   \\     
121 & 0.60	&	3.73	&	7.58	&	0.60	&	3.40	&	7.47	&	0.53	&	3.58	&	7.64   \\     
400 & 0.56	&	3.84	&	7.65	&	0.67	&	3.82	&	7.64	&	0.67	&	3.62	&	7.65   \\     
\bottomrule
\end{tabular}
\end{table}

Samples from different covariance matrices always led to rejecting the null hypothesis with the proposed statistical tests, thus leading to tests with unitary empirical test power. 
Table~\ref{tableapplication3} shows the mean value of the statistics
$$
\overline{S}(\text{A}_i,\text{A}_j)=\frac 1{5500} \sum_{k=1}^{5500} S^h_{\phi}([\widehat{\boldsymbol{\Sigma}}_i,\widehat{L}_i]_k,[\widehat{\boldsymbol{\Sigma}}_j,\widehat{L}_j]_k),
$$
where $S^h_{\phi}$ is $S_\text{S}$ or $S_\text{R}$ as given in~\eqref{teststatisticonentropy} and  $[\widehat{\boldsymbol{\Sigma}}_i,\widehat{L}_i]_k$ are the ML estimates based on generated data from population $A_i$ at the $k$th Monte Carlo replica.
The table also shows the coefficient of variation (CV) of the test statistics under alternative hypotheses:
\begin{align*}
&\operatorname{CV}(\text{A}_i,\text{A}_j)=\\
&\frac{\sqrt{\sum_{k=1}^{5500}[S^h_{\phi}([\widehat{\boldsymbol{\Sigma}}_i,\widehat{L}_i]_k,[\widehat{\boldsymbol{\Sigma}}_j,\widehat{L}_j]_k)-\overline{S}(\text{A}_i,\text{A}_j)]^2}}{\overline{S}(\text{A}_i,\text{A}_j)\sqrt{5500}}.
\end{align*}

 \begin{table}[hbt]
 \centering
 \caption{Mean and coefficient of variation under alternative hypotheses of test statistics based on Shannon and R\'enyi (of order $0.1$ and $0.8$) entropies}\label{tableapplication3}
 \begin{tabular}{rrrrrrr}\toprule
  & \multicolumn{2}{c}{$\text{A}_1$-$\text{A}_2$} & \multicolumn{2}{c}{$\text{A}_1$-$\text{A}_3$} & \multicolumn{2}{c}{$\text{A}_2$-$\text{A}_3$} \\
 \cmidrule(lr{.25em}){2-3} \cmidrule(lr{.25em}){4-5} \cmidrule(lr{.25em}){6-7}
 $N$ & $\operatorname{CV}$ & $\overline{S}$ &  $\operatorname{CV}$ & $\overline{S}$ & $\operatorname{CV}$
  & $\overline{S}$\\  \cmidrule(lr{.25em}){1-1} \cmidrule(lr{.25em}){2-3} \cmidrule(lr{.25em}){4-5} \cmidrule(lr{.25em}){6-7}
 & \multicolumn{6}{c}{Shannon entropy }\\
 9   & 20.59	&	108.16	&	13.83	&	248.53	  &	38.42	&	29.89   \\ 
 49  & 8.65	&	576.23	&	5.82	&	1329.19	  &	16.38	&	156.21  \\ 
 81  & 6.70	&	949.51	&	4.52	&	2191.84	  &	12.71	&	256.54  \\ 
 121 & 5.51	&	1418.22	&	3.71	&	3274.03	  &	10.43	&	383.17  \\ 
 400 & 3.01	&	4687.29	&	2.02	&	10817.80	&	5.72	&	1265.69 \\ 
 \cmidrule(lr{.25em}){1-1} \cmidrule(lr{.25em}){2-7}
 & \multicolumn{6}{c}{R\'enyi entropy with $\beta=0.8$ }\\
 9   & 20.56	&	106.72	&	13.72	&	245.22	  &	38.45	&	29.50   \\     
 49  & 8.63	&	569.69	&	5.76	&	1314.11	  &	16.40	&	154.43  \\     
 81  & 6.67	&	938.93	&	4.47	&	2167.39	  &	12.72	&	253.68  \\     
 121 & 5.48	&	1402.43	&	3.66	&	3237.61	  &	10.42	&	378.93  \\     
 400 & 3.00	&	4635.65	&	2.00	&	10698.61	&	5.71	&	1251.74 \\     
 \cmidrule(lr{.25em}){1-1} \cmidrule(lr{.25em}){2-7}
 & \multicolumn{6}{c}{R\'enyi entropy with $\beta=0.1$ }\\
 9 & 21.70	&	78.42  	&	14.27	&	180.12	  &	40.60	&	21.79   \\     
 49 & 9.11	&	423.29	&	6.00	&	976.32	  &	17.44	&	114.77  \\     
 81 & 7.02	&	698.47	&	4.63	&	1612.06	  &	13.50	&	188.81  \\     
 121 & 5.73	&	1043.26	&	3.77	&	2408.63	  &	11.00	&	282.06  \\     
 400 & 3.15	&	3450.84	&	2.07	&	7963.98	  &	6.03	&	931.79  \\     
 \bottomrule
 \end{tabular}
 \end{table}

The largest means $\overline{S}(\text{A}_i,\text{A}_j)$ correspond to the largest absolute values $\operatorname{abs}(|\widetilde{\boldsymbol{\Sigma}}_i|-|\widetilde{\boldsymbol{\Sigma}}_j|)$, leading to the following relations:
\begin{align}\label{inequality}
\overline{S}(\text{A}_1,\text{A}_3)>\overline{S}(\text{A}_1,\text{A}_2)>\overline{S}(\text{A}_2,\text{A}_3).
\end{align}
Shannon test statistics produced the largest test statistics when contrasting samples from different areas.

\subsection{Real Data}

\begin{figure}[htb]
\centering
\subfigure[Size $1\%$ \label{figure21}]{\includegraphics[width=\linewidth]{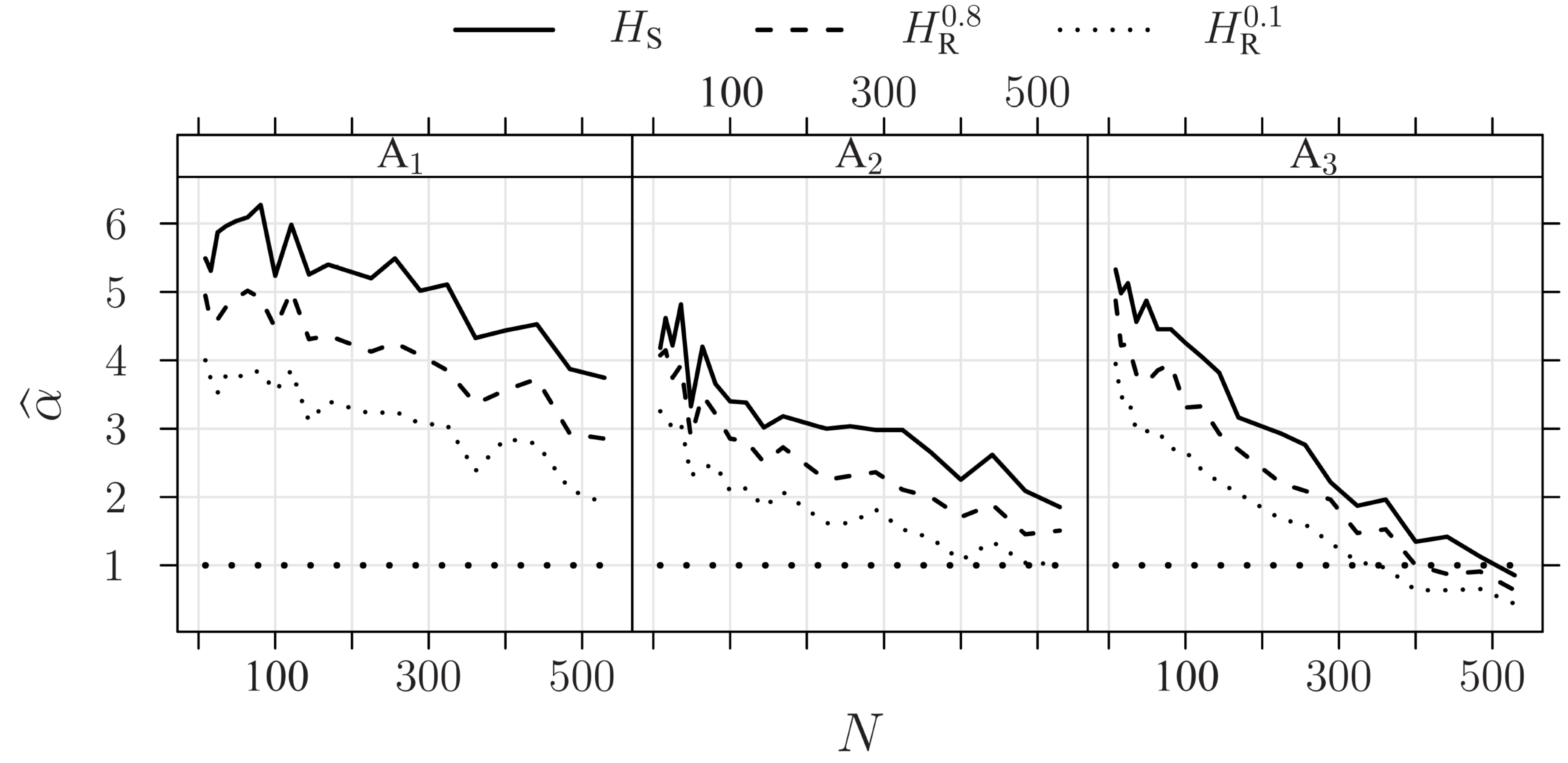}}
\subfigure[Size $5\%$ \label{figure22}]{\includegraphics[width=\linewidth]{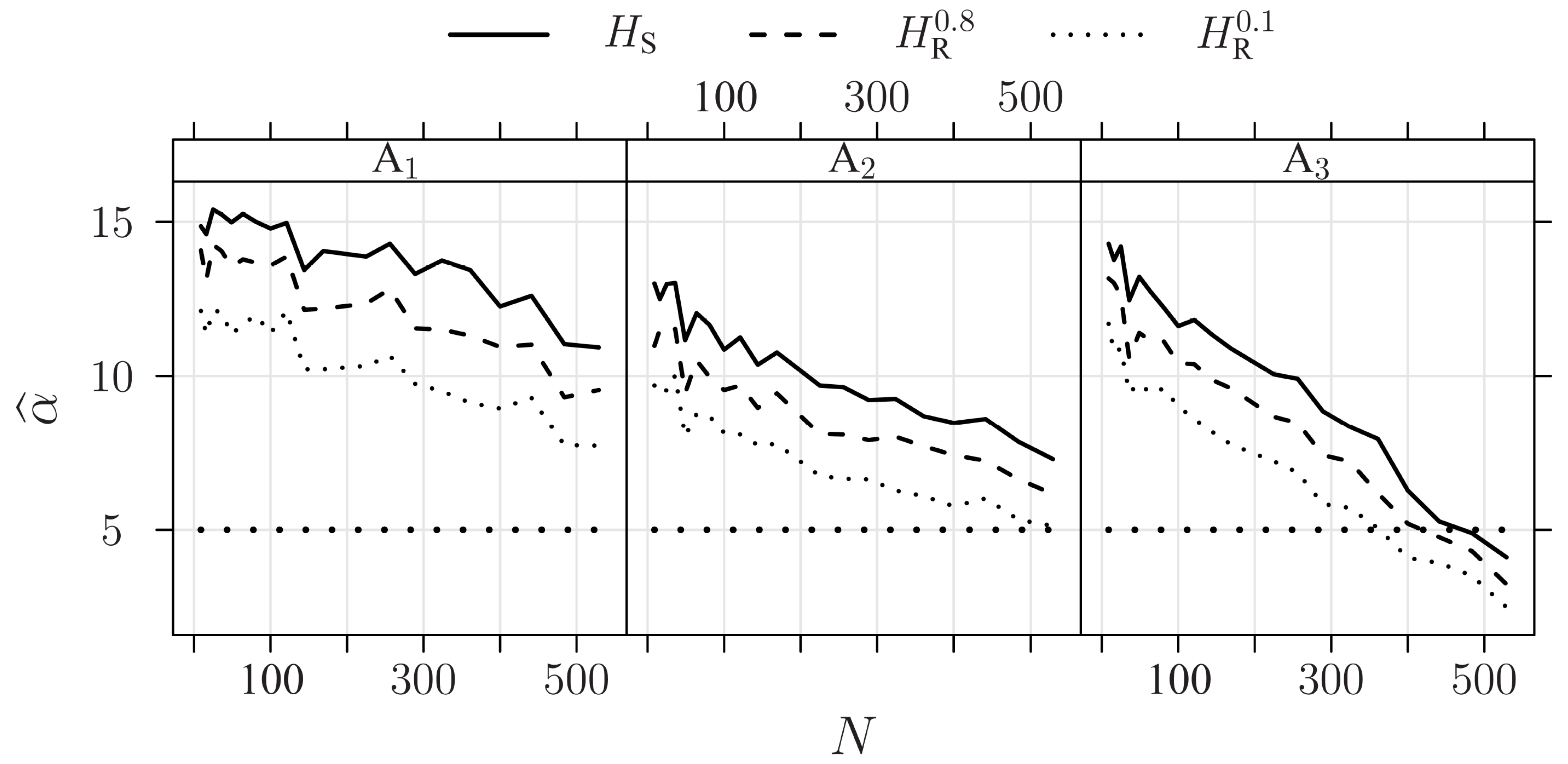}}
\subfigure[Size $10\%$ \label{figure23}]{\includegraphics[width=\linewidth]{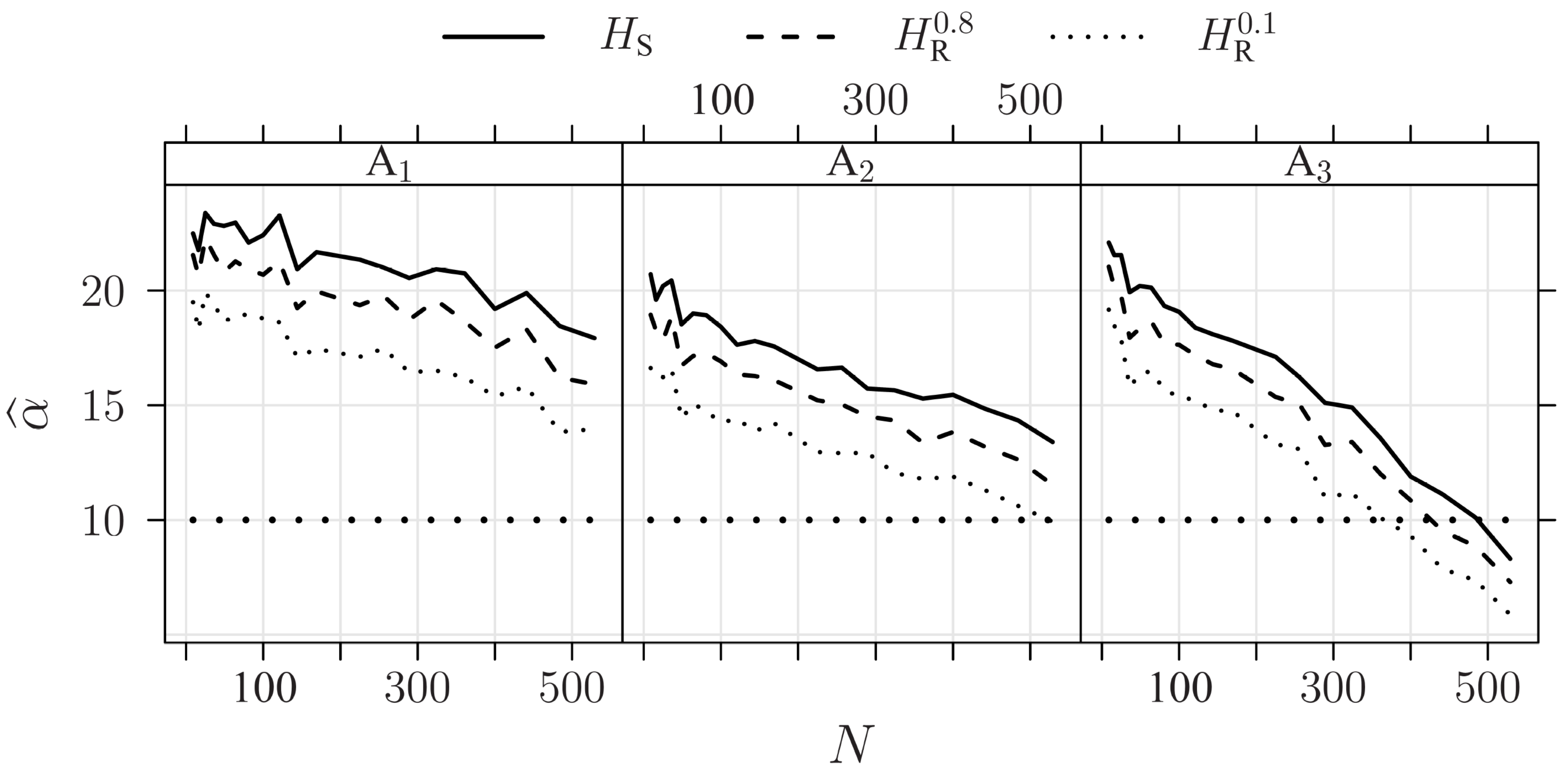}}
\caption{Empirical test size for different regions at several sample sizes and nominal levels.}
\label{figure2}
\end{figure}

Applying the methodology described above to real data, we obtained the results shown in Fig.~\ref{figure2}.
To that end, the following steps were followed: 
\begin{algorithmic}[1]
\For{$j=1,2,\ldots,5500$}\label{first_step}
	\State\label{enui} Extract two regions $\boldsymbol{U}_j$ and $\boldsymbol{V}_j$ from areas  $\text{A}_1,\text{A}_2,\text{ and }\text{A}_3$.
	\State By sampling without replacement, generate two $N$-point vectors $\boldsymbol{u}^{(j)}$ and $\boldsymbol{v}^{(j)}$ from $\boldsymbol{U}_j$ and $\boldsymbol{V}_j$, respectively, granting $\boldsymbol{u}^{(i)}\neq\boldsymbol{u}^{(j)}$ and $\boldsymbol{v}^{(i)}\neq\boldsymbol{v}^{(j)}$ for $i\neq j$.
	\State Estimate the parameter vectors $\widehat{\boldsymbol{\theta}}^{(j)}_1$ and $\widehat{\boldsymbol{\theta}}^{(j)}_2$ based on $\boldsymbol{u}^{(j)}$ and $\boldsymbol{v}^{(j)}$, respectively.
	\State\label{enuf} Compute the decision from Proposition~\ref{p1} for $\alpha =\{1\%,5\%,10\%\}$.
\EndFor
\State\label{last_step} Let $T$ be the number of times that the null hypothesis is rejected. 
Calculate the empirical test size ($\widehat{\alpha}_{1-\alpha}$) and power ($1-\widehat{\beta}_{1-\alpha}$) at level $\alpha$ as
$$
\left\{ \begin{array}{c}
\widehat{\alpha}_{1-\alpha}=T/5500,\text{ if }\boldsymbol{V}_j=\boldsymbol{U}_j, \\
1-\widehat{\beta}_{1-\alpha}=T/5500,\text{ if }\boldsymbol{V}_j\neq\boldsymbol{U}_j,
\end{array} \right.
$$
\end{algorithmic}
where $\widehat{\beta}_{1-\alpha}$ is the Type~II error rate; i.e., the estimate for the probability of not rejection of $\mathcal H_0$ when the null hypothesis is false.  
We considered $N$ the number of observations in squared windows of size $\{3\times 3, 4\times 4,\ldots,23\times 23\}$.

Fig.~\ref{figure2} shows the empirical test sizes, with the nominal test size $\alpha$ in dotted line.
The behavior of $\widehat\alpha_{1-\alpha}$ is consistent across the three values of $\alpha$: the empirical test size is larger in Shannon  than in  R\'enyi $\beta=0.8$ which, in turn, is larger than R\'enyi $\beta=0.1$. 
The fastest convergence to $\alpha$ with respect to the sample size is consistently observed in area $\text{A}_3$. 
No test attains the nominal size in area $\text{A}_1$.
This is probably due to its pronounced brightness and roughness, as noted in Table~\ref{tableapplication1}.

\begin{figure}[htb]
\centering
\subfigure[EMISAR Image\label{powerr1}]{\includegraphics[width=.48\linewidth]{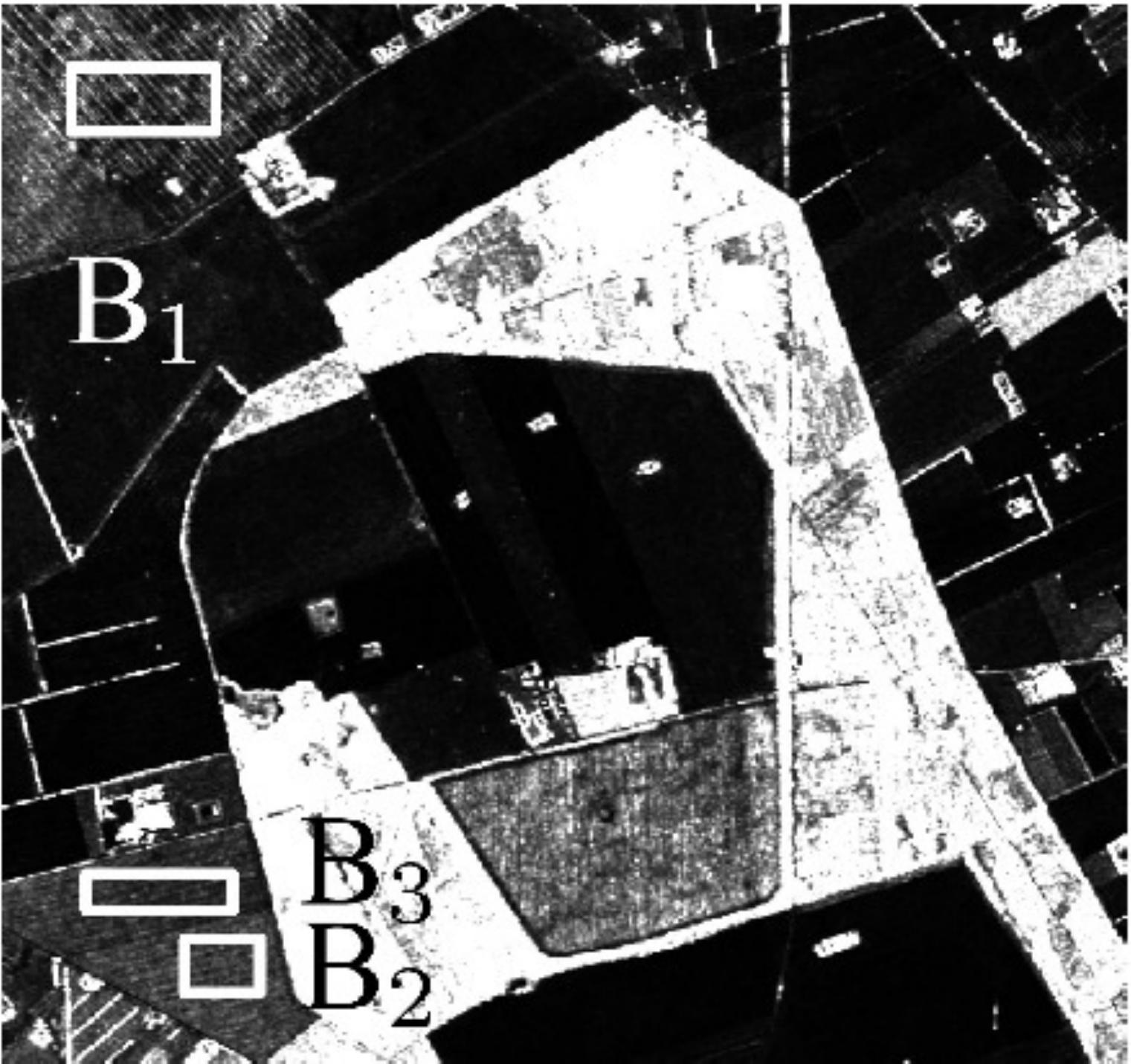}}
\subfigure[HH Channel\label{powerr2}]{\includegraphics[width=.48\linewidth]{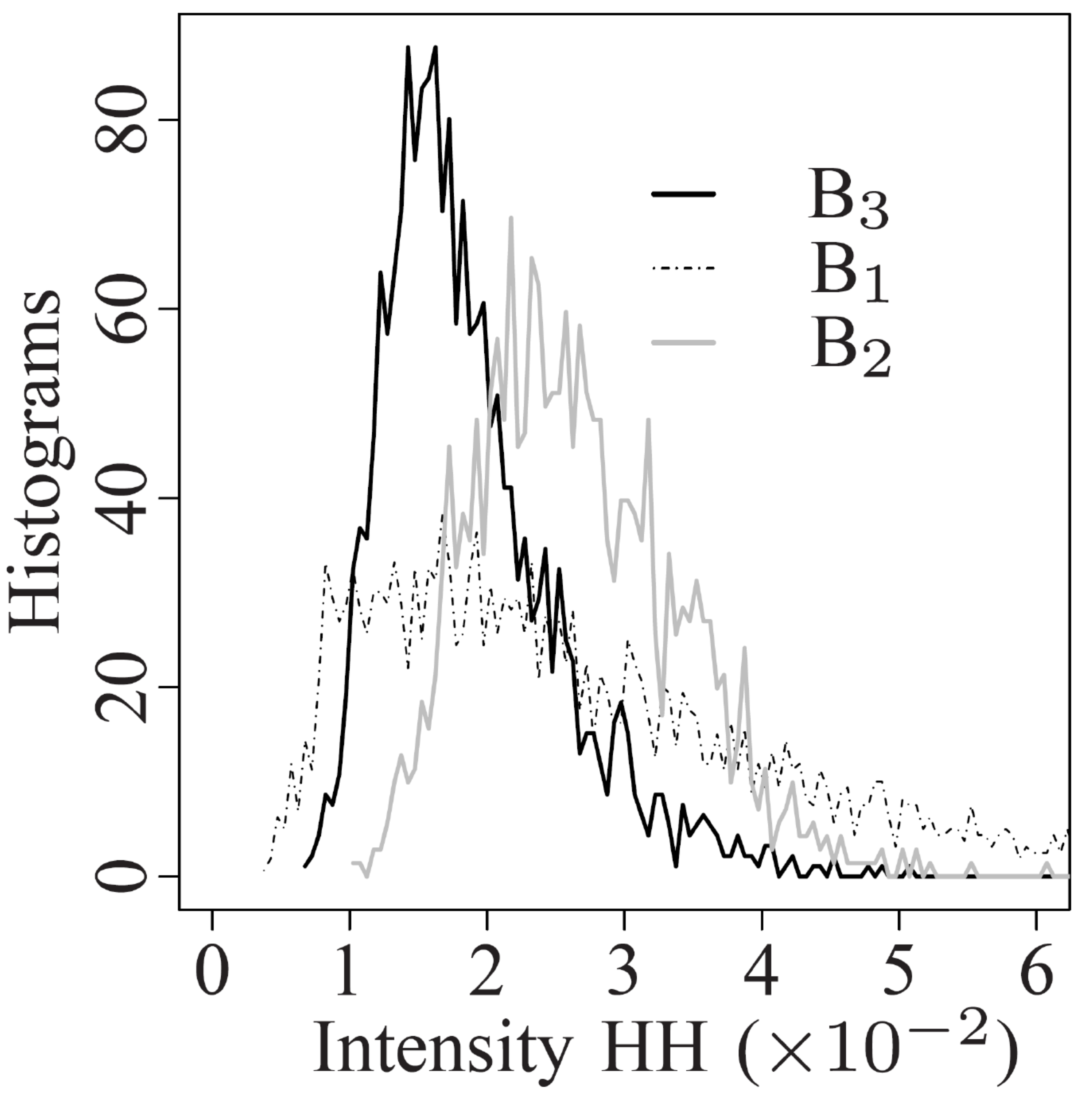}}
\subfigure[HV Channel\label{powerr3}]{\includegraphics[width=.48\linewidth]{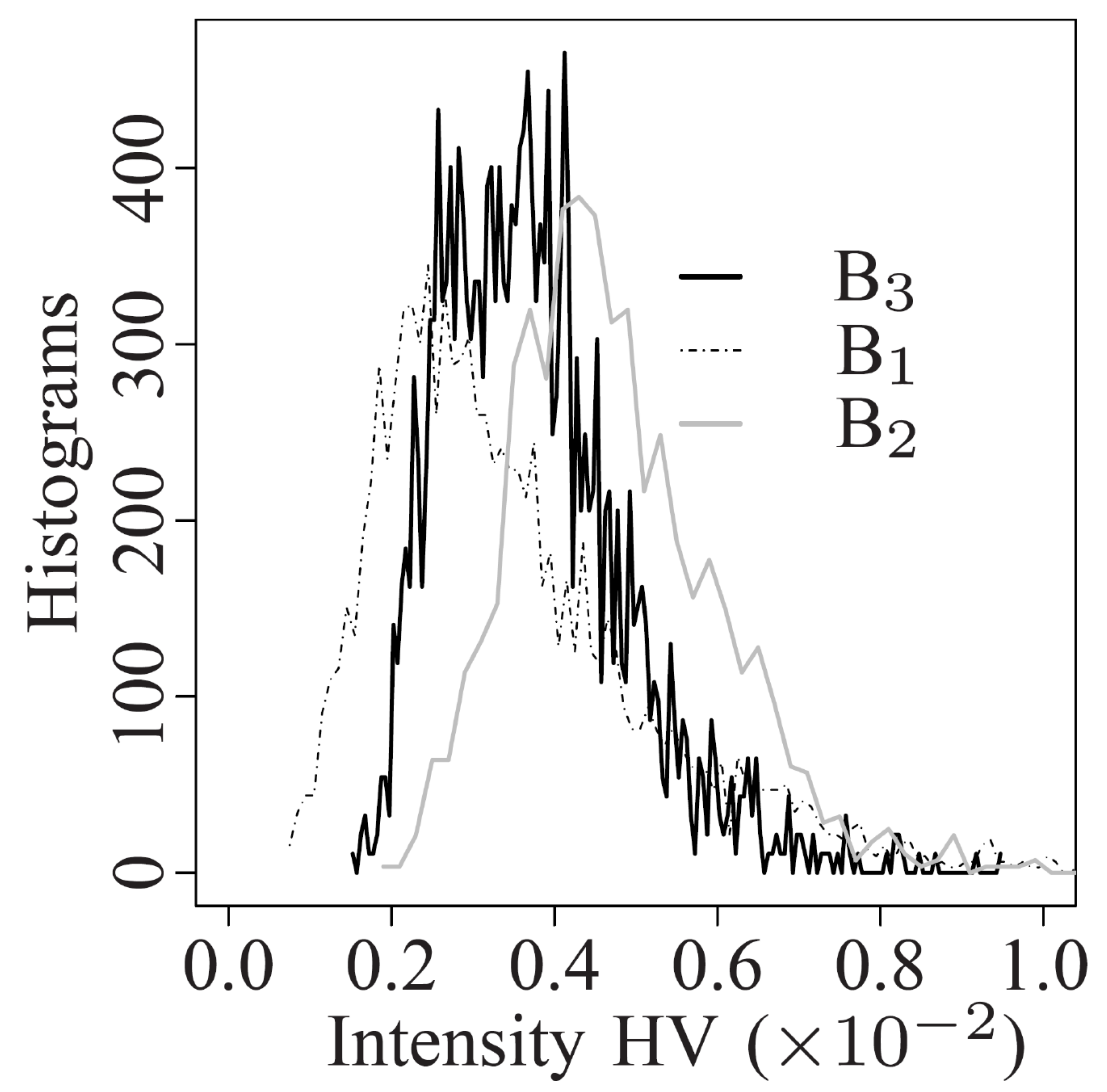}}
\subfigure[VV Channel\label{powerr4}]{\includegraphics[width=.48\linewidth]{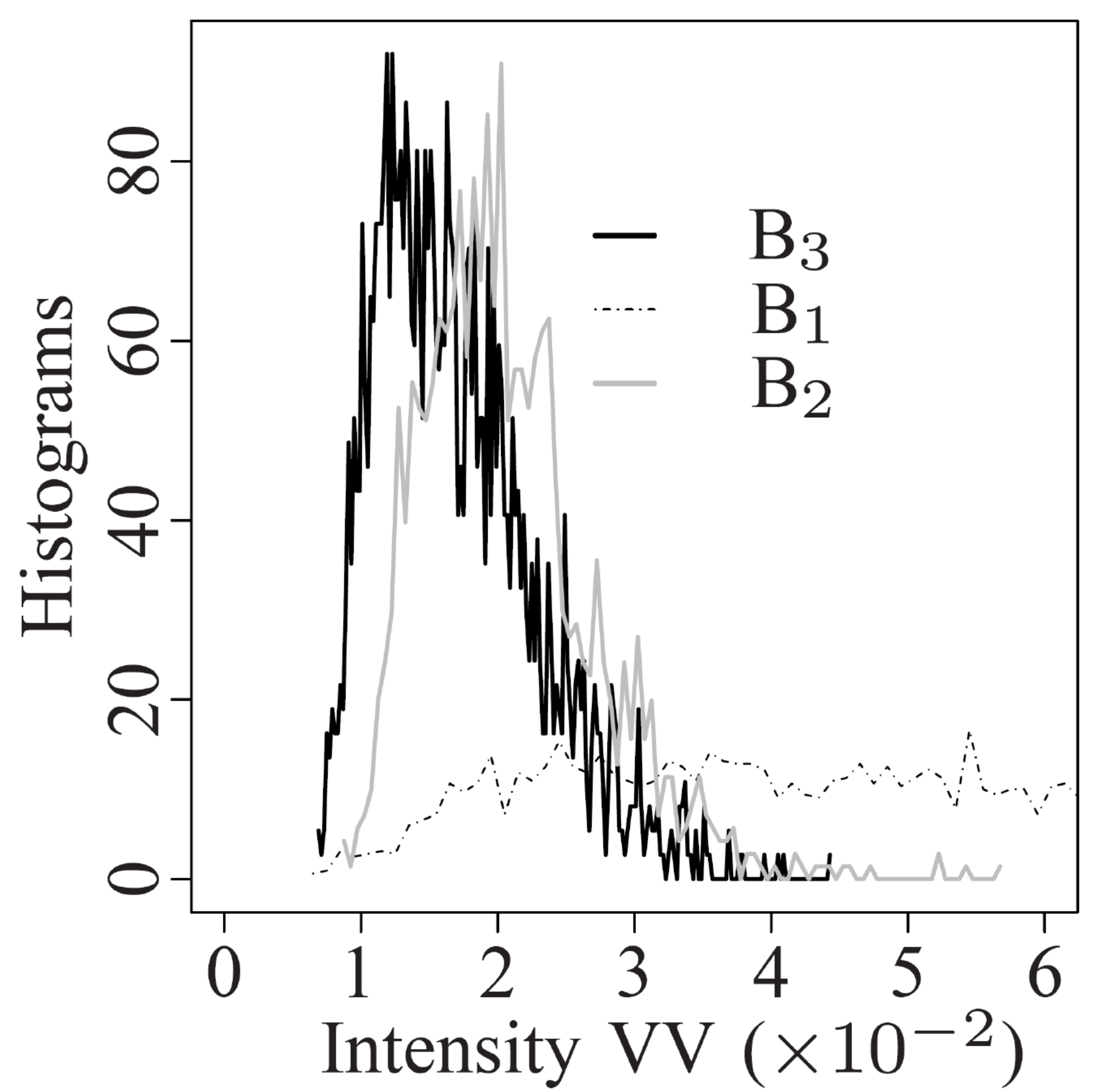}}
\caption{PolSAR image, selected regions and their empirical densities in all polarizations.} 
\label{passa0}
\end{figure}

In the following we select visually similar regions, but with distinct statistical properties.
To that end, Fig.~\ref{powerr1} shows an EMISAR image of Foulum (Denmark) obtained with eight nominal looks over agricultural areas and sub \unit[10]{m} resolution; three regions were selected for our study. 
Table~\ref{tableapplication23} presents the sample sizes and the ML estimates, while Figs.~\ref{powerr2}, \ref{powerr3} and~\ref{powerr4} show empirical densities of different areas and channels. 
The empirical densities of B$_2$ and B$_3$ are the closest for all polarization channels, as well as their estimates of $L$.

\begin{table}
\centering 
\caption{ML estimates for the samples from Figure~\protect{\normalfont{\ref{powerr1}}}}\label{tableapplication23}
\begin{tabular}{cccc}
\toprule                                                           
Regions & $N$   & $|\widehat{\boldsymbol{\Sigma}}|$ &  $\widehat{L}$ \\ \midrule
$\text{B}_1$ & 3192   &  $1.609\times 10^{-5}$  &  6.925 \\
$\text{B}_2$ & 1408   &  $1.112\times 10^{-6}$  &  11.937  \\
$\text{B}_3$ & 1848   &  $5.814\times 10^{-7}$  &  10.752  \\
\bottomrule
\end{tabular}
\end{table}

\begin{figure}[htb]
\centering
\includegraphics[width=.48\linewidth]{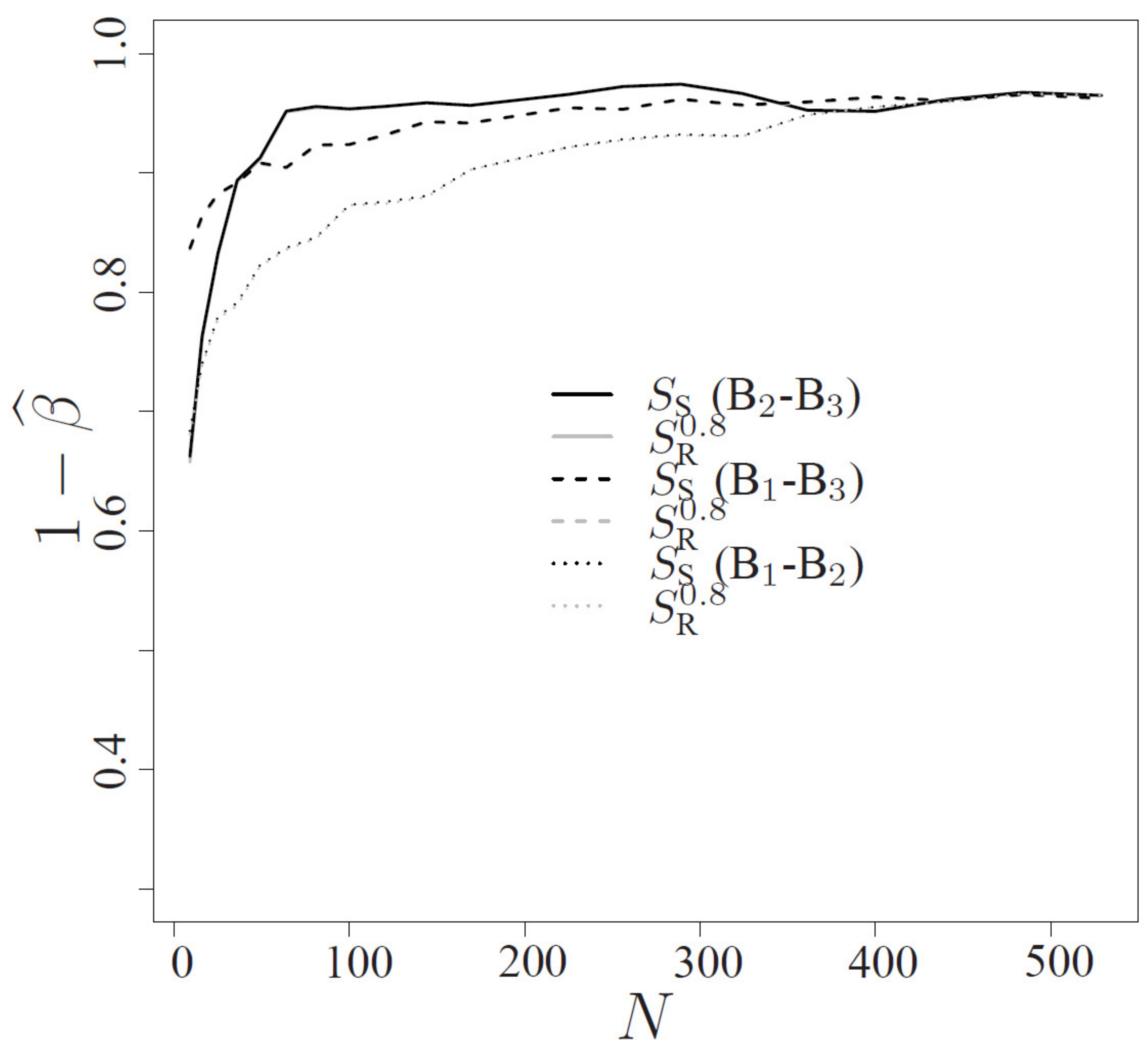}
\caption{Empirical test power at the level $10\%$.} 
\label{addpower}
\end{figure}

We quantify the discrimination between selected regions by means of $S_\text{S}$ and $S_\text{R}^{0.1}$ performing steps~\ref{first_step}--\ref{last_step} with the Foulum samples.
Since R\'enyi entropy converges to Shannon when the order tends to $1$, we omit the analysis of $S_\text{R}^{0.8}$ because it is similar to $S_\text{S}$, as illustrated in Fig.~\ref{addpower}.

\begin{figure}[htb]
\centering
\subfigure[1\%\label{powerr5}]{\includegraphics[width=.48\linewidth]{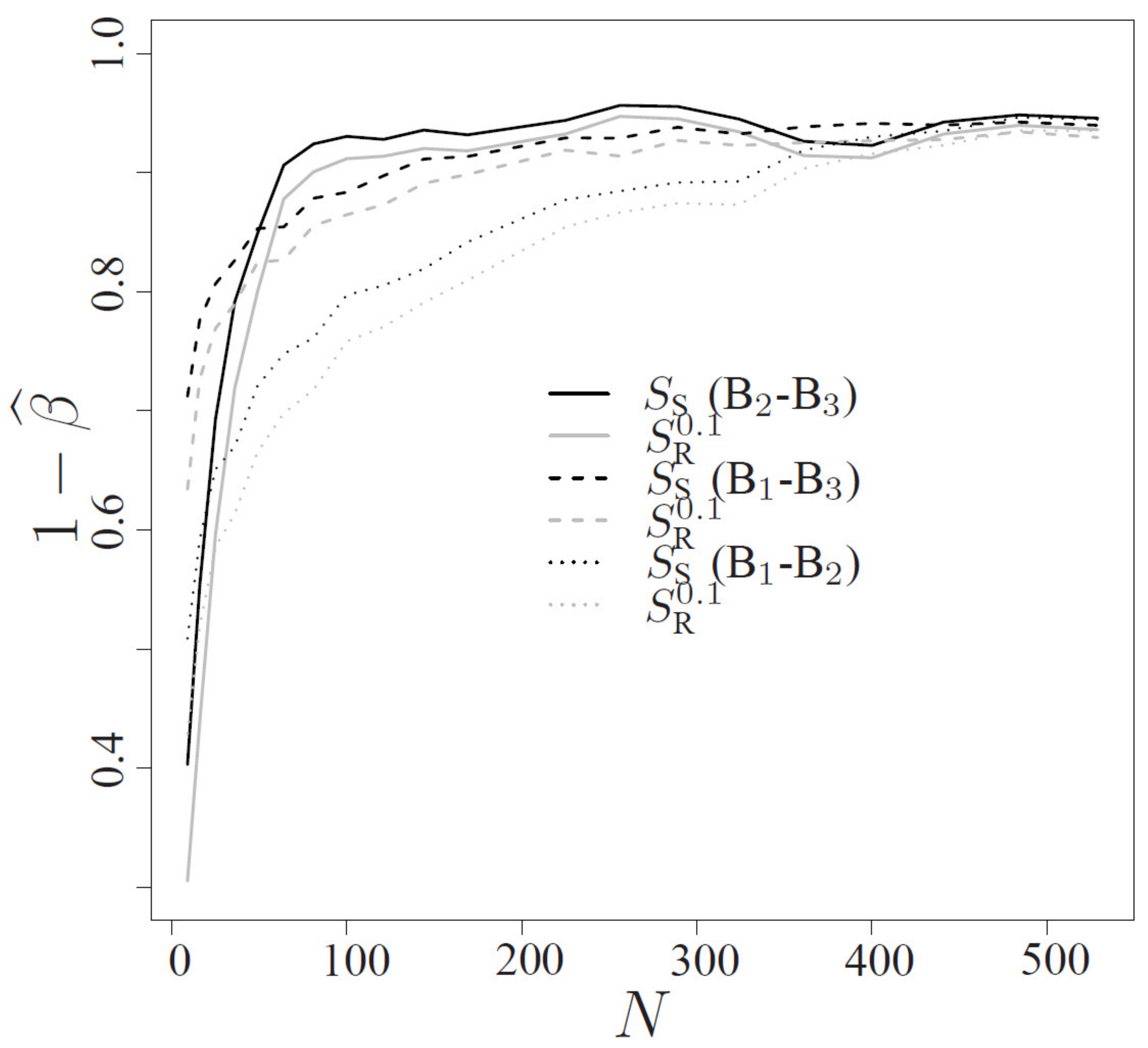}}
\subfigure[5\%\label{powerr6}]{\includegraphics[width=.48\linewidth]{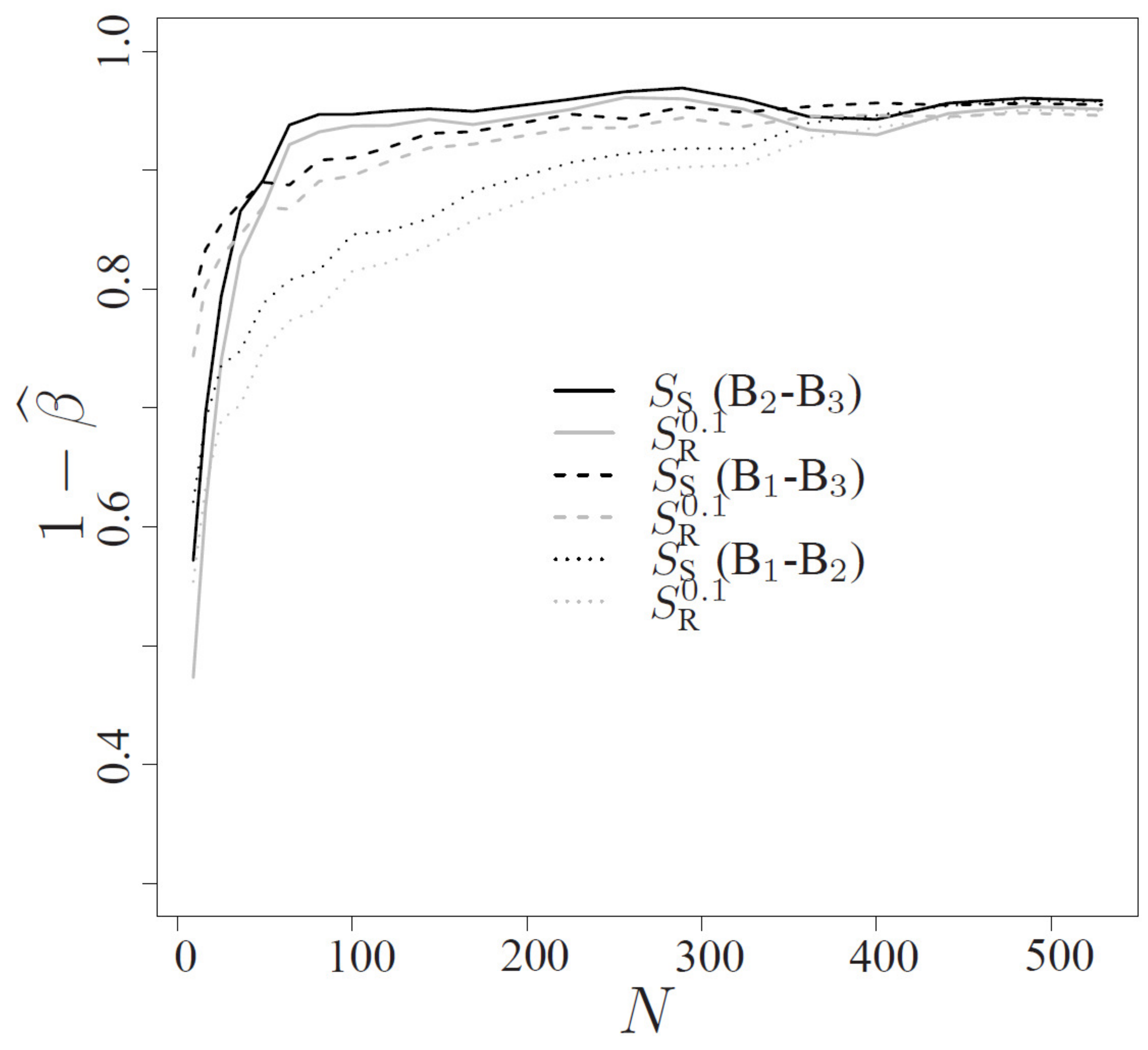}}
\subfigure[10\%\label{powerr7}]{\includegraphics[width=.48\linewidth]{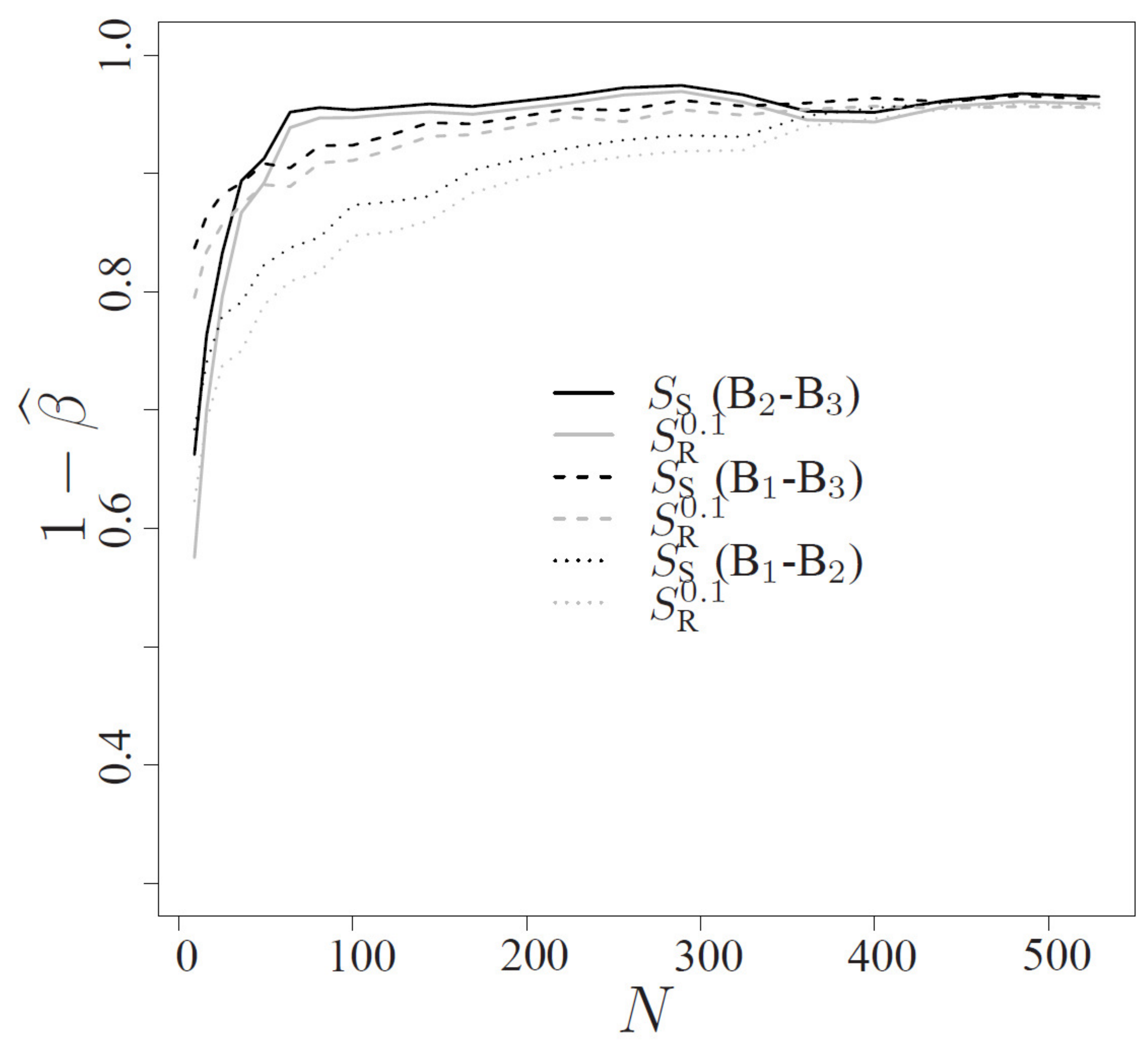}}
\caption{Empirical test power for different regions at several sample sizes and nominal levels; a good test has high power in these cases.} 
\label{passa1}
\end{figure}

Figs.~\ref{powerr5}, \ref{powerr6} and~\ref{powerr7} exhibit the empirical test power. 
In general terms, the estimated power increase quickly when sample sizes increase.  
The test based on the Shannon entropy outperforms the one based on R\'enyi's. 

These results suggest that the test statistics in terms of the Shannon entropy is a more appropriate tool for change and boundary detection in PolSAR images than the measure based on the R\'enyi entropy.

\section{Conclusions}\label{entropy:conclusion} 

In this paper, the closed expressions for Shannon, R\'enyi, and Tsallis entropies have been derived for data distributed according with the scaled complex Wishart distribution. 
The variances of the two first entropies were also derived when parameters are replaced by ML estimators, leading to the derivation of two test statistics with known asymptotic distributions. 
The statistics based on the Tsallis entropy were not derived since its variance could not be expressed in closed form.

New hypothesis tests and confidence intervals based on the ($h,\phi$)-family of entropies were derived for the scaled complex Wishart distribution.
These new statistical tools provide means for contrasting $r\geq2$ samples and verifying if they come from the same distribution.
Such tools were derived by obtaining orthogonal maximum likelihood estimators for the scaled complex Wishart law, and by characterizing their asymptotic behavior.

Monte Carlo experiments were employed for quantifying the performance of the proposed hypotheses tests.
The results provided evidence that the statistic based on the Shannon entropy is the most efficient in terms of empirical test size and power. 

An application to actual data was also considered in order to assess the performance of the proposed hypothesis tests.
The empirical test sizes observed with real data are relatively high when dealing with small samples, but they decrease as the sample size increases.
As expected, the tests presented the worst results in the most heterogeneous situations.
Nevertheless, they perform correctly when the sample size is increased.

The hypothesis test based on the Shannon entropy presented the smallest empirical test size.
All test statistics detected differences among different regions at the specified levels.
This is important for PolSAR image analysis, such as in boundary detection~\cite{PolarimetricSegmentationBSplinesMSSP} and change detection~\cite{IngladaMercier2007}. 

As an overall conclusion, the Shannon entropy can be safely used for discriminating areas in PolSAR imagery.
Moreover, the Shannon entropy it is not challenged in terms of simplicity either, which consolidates its position as the preferred entropy measure.
However, care must be taken when small samples are analyzed. 
Indeed, in this case, the proposed tests are prone to classifying regions of similar natures as distinct, i.e., to incurring in Type~I error. 
In practice, however, this issue is not severe, since PolSAR image processing often handles data with a large number of pixels.

Further research will consider models which include heterogeneity~\cite{FreitasFreryCorreia:Environmetrics:03,FreryCorreiaFreitas:ClassifMultifrequency:IEEE:2007,
PolarimetricSegmentationBSplinesMSSP}, robust, improved and nonparametric inference~\cite{AllendeFreryetal:JSCS:05,NonparametricEdgeDetectionSpeckledImagery,SilvaCribariFrery:ImprovedLikelihood:Environmetrics,
VasconcellosFrerySilva:CompStat}, and small samples issues~\cite{FreryCribariSouza:JASP:04}.

\bibliographystyle{IEEEtran}
\bibliography{BibEntropyII} 

\newpage
\begin{IEEEbiography}[{\includegraphics[width=1in]{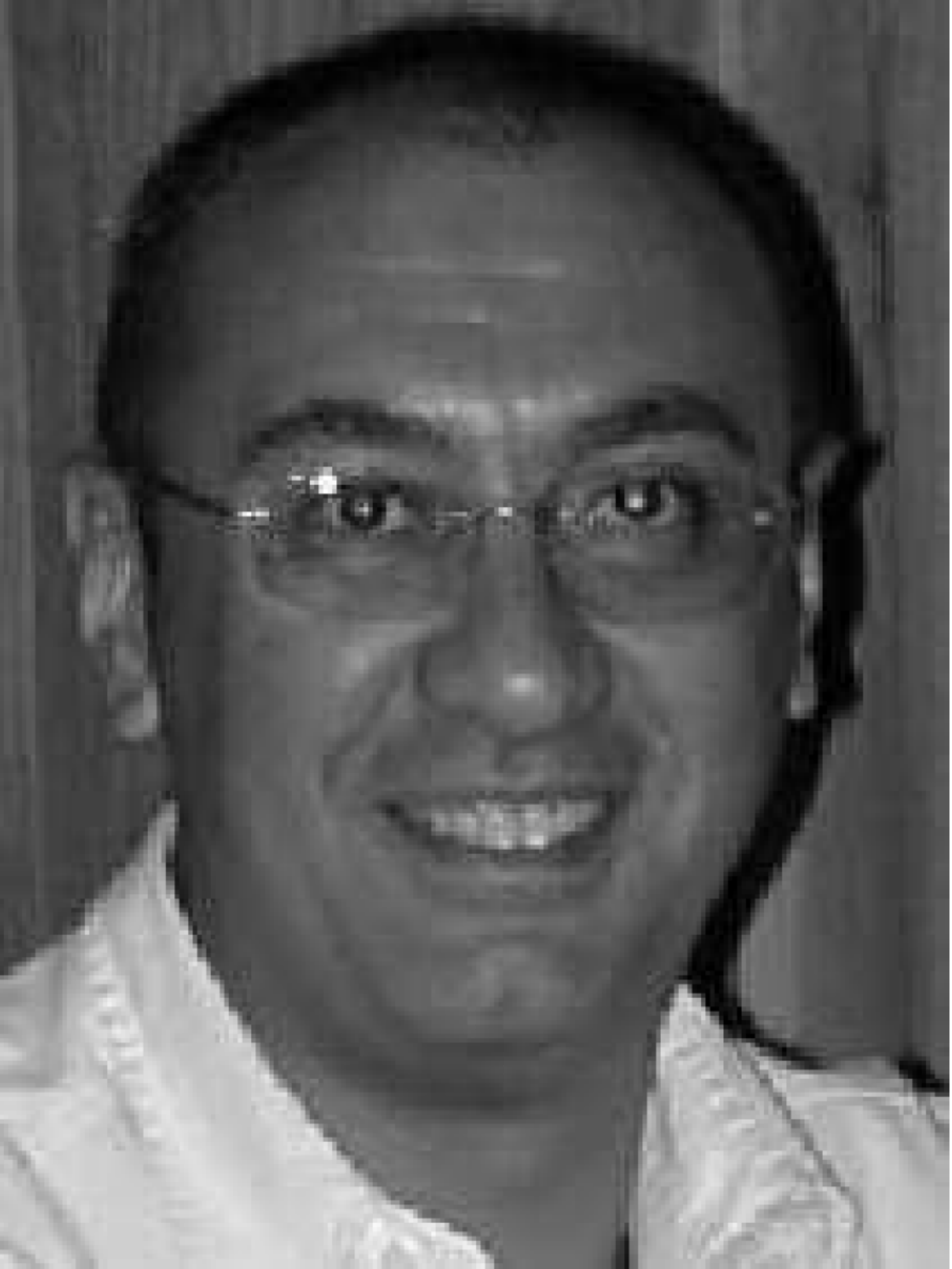}}]{Alejandro C.\ Frery}
graduated in Electronic and Electrical Engineering from the Universidad de Mendoza, Argentina.
His M.Sc. degree was in Applied Mathematics (Statistics) from the Instituto de Matem\'atica Pura e Aplicada (Rio de Janeiro) and his Ph.D. degree was in Applied Computing from the Instituto Nacional de Pesquisas Espaciais (S\~ao Jos\'e dos Campos, Brazil).
He is currently with the Instituto de Computa\c c\~ao, Universidade Federal de Alagoas, Macei\'o, Brazil.
His research interests are statistical computing and stochastic modelling.
\end{IEEEbiography}

\begin{IEEEbiography}[{\includegraphics[width=1in]{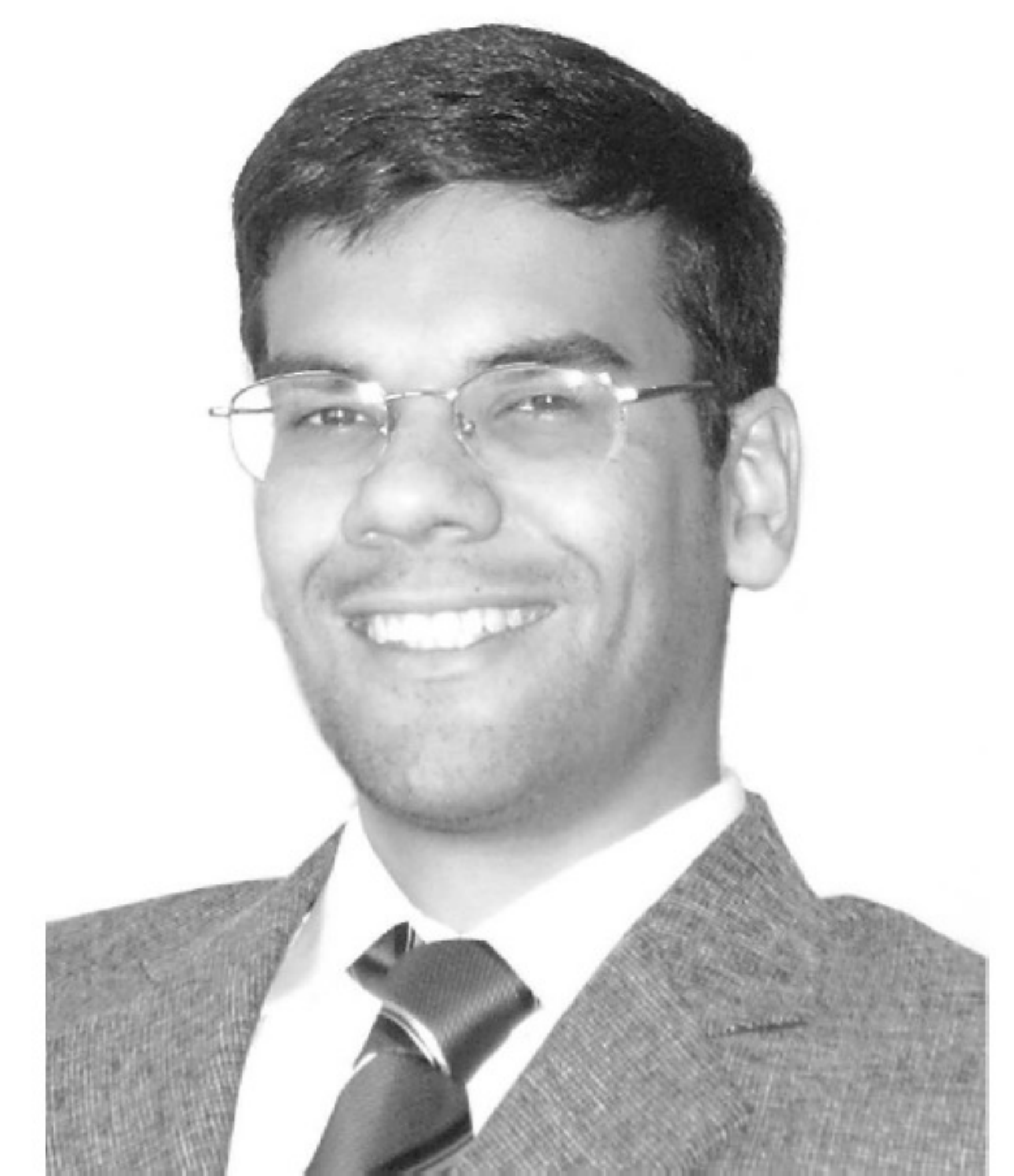}}]{Renato J.\ Cintra}
earned his B.Sc., M.Sc., and D.Sc. degrees
in Electrical Engineering from
Universidade Federal de Pernambuco,
Brazil, in 1999, 2001, and 2005, respectively.
In 2005,
he joined the Department of Statistics at UFPE.
During 2008-2009,
he worked at the University of Calgary, Canada,
as a visiting research fellow.
He is also a graduate faculty member of the
Department of Electrical and Computer Engineering,
University of Akron, OH.
His long term topics of research include
theory and methods for digital signal processing,
communications systems, and applied mathematics.
\end{IEEEbiography}

\begin{IEEEbiography}[{\includegraphics[width=1in]{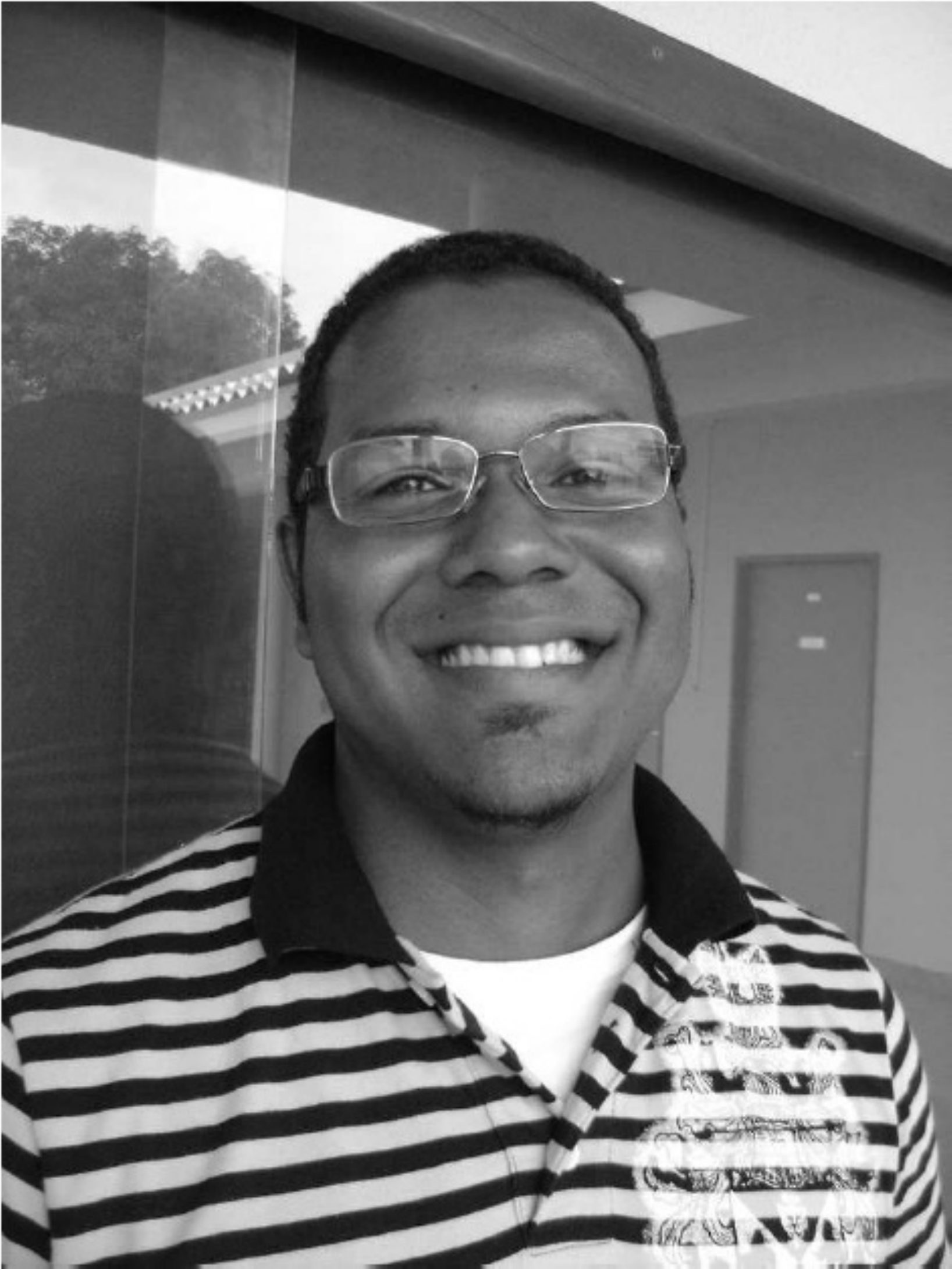}}]{Abra\~ao D.\ C.\ Nascimento}
holds B.Sc.\, M.Sc.\, and D.Sc. degrees in Statistics from Universidade Federal de Pernambuco (UFPE), Brazil, in 2005, 2007, and 2012, respectively.
In 2012, he joined the Department of Statistics at UFPE as Substitute Professor.
His research interests are statistical information theory, inference on random matrices, and asymptotic theory.
\end{IEEEbiography}

\vfill

\end{document}